\pdfoutput=1

\documentclass[11pt,twoside,a4paper,cmspaper,final,collab]{cms-tdr}
\def\svnVersion{272775}\def\cmsCernNo{2013-037}\def\cmsCernDate{\today}\def\cmsMessage{Published in Physical Review D as \href{http://dx.doi.org/10.1103/PhysRevD.90.112013}{\doi{10.1103/PhysRevD.90.112013.}}}

\begin{document}\cmsNoteHeader{HIG-13-025}

\hyphenation{had-ron-i-za-tion}
\hyphenation{cal-or-i-me-ter}
\hyphenation{de-vices}
\RCS$HeadURL: svn+ssh://svn.cern.ch/reps/tdr2/papers/HIG-13-025/trunk/HIG-13-025.tex $
\RCS$Id: HIG-13-025.tex 267367 2014-11-12 11:45:40Z alverson $

\newlength\cmsFigWidth
\ifthenelse{\boolean{cms@external}}{\setlength\cmsFigWidth{0.95\columnwidth}}{\setlength\cmsFigWidth{0.6\textwidth}}
\ifthenelse{\boolean{cms@external}}{\providecommand{\cmsLeft}{top}}{\providecommand{\cmsLeft}{left}}
\ifthenelse{\boolean{cms@external}}{\providecommand{\cmsRight}{bottom}}{\providecommand{\cmsRight}{right}}
\ifthenelse{\boolean{cms@external}}{\providecommand{\cmsLLeft}{Top}}{\providecommand{\cmsLLeft}{Left}}
\ifthenelse{\boolean{cms@external}}{\providecommand{\cmsRRight}{Bottom}}{\providecommand{\cmsRRight}{Right}}
\ifthenelse{\boolean{cms@external}}{\providecommand{\NA}{\ensuremath{\cdots}}}{\providecommand{\NA}{\text{---}}}
\ifthenelse{\boolean{cms@external}}{\providecommand{\CL}{C.L.\xspace}}{\providecommand{\CL}{CL\xspace}}

\newcommand{\procLumi}{19.5\fbinv}

\providecommand{\tauh}{\ensuremath{\Pgt_\mathrm{h}}\xspace}
\providecommand{\PA}{\ensuremath{\cmsSymbolFace{A}}\xspace}
\providecommand{\Ph}{\ensuremath{\cmsSymbolFace{h}}\xspace}

\cmsNoteHeader{HIG-13-025}
\title{\texorpdfstring{Searches for heavy Higgs bosons in two-Higgs-doublet models and for $\cPqt \to \cPqc \Ph$ decay using multilepton and diphoton final states in $\Pp\Pp$ collisions at $8\TeV$}
{Searches for heavy Higgs bosons in two-Higgs-doublet models and for t -> ch decay using multilepton and diphoton final states in pp collisions at 8 TeV}
}

\date{\today}

\abstract{Searches are presented for heavy scalar ($\PH$) and pseudoscalar
($\PA$) Higgs bosons posited in the two doublet model (2HDM) extensions
of the standard model (SM).
These searches are based on a data sample of
$\Pp \Pp$ collisions collected with the CMS experiment at the LHC at a center-of-mass energy
of $\sqrt{s} = 8\TeV$ and corresponding to an integrated luminosity of 19.5\fbinv.
The decays $\PH \to \Ph \Ph$ and $\PA \to \cPZ \Ph$, where $\Ph$ denotes an
SM-like Higgs boson, lead to events with three or more isolated charged leptons or with a
photon pair accompanied by one or more isolated leptons.
The search results are presented in terms of the $\PH$ and $\PA$
production cross sections times branching fractions
and are further interpreted in terms of 2HDM parameters.
We place 95\% \CL cross section upper limits of  approximately 7\unit{pb} on $\sigma \mathcal{B}$
for $\PH \to \Ph \Ph$ and 2\unit{pb} for $\PA \to \cPZ \Ph$.
Also presented are the results of a search for the rare decay of the top quark that results
in a charm quark and an SM Higgs boson, $\cPqt \to \cPqc \Ph$, the existence of which
would indicate a nonzero flavor-changing Yukawa coupling of the top quark to the Higgs boson.
We place a 95\% \CL upper limit of 0.56\% on $\mathcal{B}(\cPqt \to \cPqc \Ph)$.
}

\hypersetup{%
pdfauthor={CMS Collaboration},%
pdftitle={Searches for heavy Higgs bosons in two-Higgs-doublet models and for t to ch decay using multilepton and diphoton final states in pp collisions at 8 TeV},%
pdfsubject={CMS},%
pdfkeywords={Higgs, BSM, 2HDM, top}}

\maketitle

\section{Introduction}
\label{sec:intro}

The standard model (SM) has an outstanding record of consistency
with experimental observations. It is not a complete theory, however, and since the
recent discovery of a Higgs boson~\cite{Aad:2012tfa,Chatrchyan:2012ufa,Chatrchyan:2013lba},
attaining a better
understanding of the mechanism responsible for
electroweak symmetry breaking (EWSB) has become a central goal in particle physics.
The experimental directions to pursue this goal include improved
characterization of the Higgs boson properties, searches for new particles
such as the members of an extended Higgs sector or the partners of the
known elementary particles predicted by supersymmetric models,
and searches for unusual processes such as rare decays of the top quark.
Since the Higgs boson plays a critical role in EWSB,
searches and studies of decays with the Higgs boson in the final state have
become particularly attractive.

In many extensions of the SM,
the Higgs sector includes two scalar
doublets~\cite{Branco:2011iw}. The two Higgs doublet
model~(2HDM)~\cite{Craig:2012vn} is a specific example of such an SM extension.
In this model five physical Higgs sector
particles survive EWSB: two neutral CP-even scalars ($\Ph$, $\PH$), one
neutral CP-odd pseudoscalar ($\PA$), and two charged scalars
($\PH^{+}$, $\PH^{-}$)~\cite{Craig:2012pu}. For masses at or below the 1\TeV scale
these particles can be produced at the LHC.
Both the heavy scalar $\PH$ and pseudoscalar $\PA$ can decay into electroweak bosons,
including the recently discovered Higgs boson. The branching fractions of $\PH$ and $\PA$ into
final states containing one or more Higgs bosons $\Ph$ often dominate when kinematically
accessible.
For heavy scalars with masses below the top pair production threshold,
the $\PH \to \Ph \Ph$ and $\PA \to \cPZ \Ph$ decays typically dominate over competing
Yukawa decays to bottom quarks, while for heavy scalars with masses above the top pair
production threshold,
these decays are often comparable in rate to
decays into top pairs and are potentially more distinctive.

We describe a search for two members of the extended Higgs sector, $\PH$ and
$\PA$, via their decays $\PH \to \Ph \Ph$ and $\PA \to \cPZ \Ph$, where $\Ph$ denotes the recently
discovered SM-like Higgs boson~\cite{Aad:2012tfa,Chatrchyan:2012ufa,Chatrchyan:2013lba}.
The final states used in this search
consist of three or more charged leptons or a resonant photon pair
accompanied by at least one charged lepton. (In the remainder of this paper,
``lepton'' refers to a charged lepton, $\Pe$, $\Pgm$, or hadronic decay of the
$\tau$-lepton, $\tauh$).
The $\PH \to \Ph \Ph$ and $\PA \to \cPZ \Ph$
decays can yield multileptonic final states when $\Ph$
decays to $\PW \PW^*$, $\cPZ \cPZ^*$, or $\Pgt \Pgt$.
Similarly, the resonant
decay $\Ph \to \gamma \gamma$
can provide a final state that contains a photon pair and one or more
leptons from the decay of the other daughter particle.

Using the same dataset and technique, we also investigate the process
$\cPqt \to \cPqc \Ph$, namely the flavor changing rare decay of
the top quark to a Higgs boson accompanied by a charm quark in the
$\ttbar \to (\cPqb \PW)(\cPqc \Ph)$ decay.
The $\cPqt \to \cPqc \Ph$ process can occur
at an observable rate for some parameters of the 2HDM
~\cite{Craig:2012vj}. Depending on how the $\Ph$ boson
and $\cPqt$
quark decay, both the multilepton and the lepton+diphoton final states can be
produced.
Both ATLAS~\cite{Aad:2014dya} and CMS~\cite{Chatrchyan:2014aea} have searched for this
process using complementary techniques. The CMS upper limit for the
branching fraction of 1.3\% at 95\% Confidence Level (\CL) comes from
an inclusive multilepton search that uses the dataset analyzed here.
We describe here a $\cPqt \to \cPqc \Ph$
search using lepton+diphoton events and combine the results of
the previously reported multilepton search with the present
lepton+diphoton search. This combination results in a considerable
improvement in the
$\cPqt \to \cPqc \Ph$
search sensitivity.

In this paper, we first briefly describe the CMS
detector,
data collection, and the detector simulation
scheme in Section~\ref{detector}.  We then describe
in Section~\ref{objects}
the selection of events that are relevant
for the search signatures followed by the event classification
in Section~\ref{strategy}, which calls
for the data sample to be subdivided in a number of mutually exclusive
channels based on the number and flavor of leptons, the number of hadronically
decaying $\tau$ leptons, photons, the tagged flavors of the jets, as well as
the amount of missing transverse
energy ($\MET$).  A description of the SM background estimation
in Section~\ref{bkgnd}
precedes the channel-by-channel comparison of the observed number
of events with
the background estimation
in Section~\ref{obs}. We next interpret
in Section~\ref{interpr}
these observations in
terms of the standalone production and decay rates
for $\PH$ and $\PA$. Since these rates follow from the
parameters of the 2HDM, we reexpress these results in terms of
the appropriate 2HDM parameters.  Finally, we selectively redeploy the
$\PH$ and $\PA$ analysis procedure to search for the rare $\cPqt \to \cPqc \Ph$ decay.

The multilepton component of this analysis closely
follows the previously mentioned CMS inclusive multilepton
analysis~\cite{Chatrchyan:2014aea}. In particular, the lepton
reconstruction, SM background estimation procedures
as well as the dataset used are identical in the two analyses
and are therefore described minimally here.

\section{Detector, data collection, and simulation}
\label{detector}

The central feature of the CMS detector is a superconducting solenoidal magnet of field
strength 3.8\unit{T}. Within the field volume are a silicon pixel and strip tracker, a lead tungstate
crystal calorimeter, and a brass-and-scintillator hadron calorimeter. The tracking detector
covers the pseudorapidity region $\abs{\eta} < 2.5$ and the calorimeters $\abs{\eta} < 3.0$.
Muon detectors based on gas-ionization detectors lie outside the solenoid, covering $\abs{\eta} < 2.4$.
A steel-and-quartz-fiber forward calorimeter provides additional coverage
between $3 < \abs{\eta} < 5.0$.
A detailed description of the detector as well as a description of the
coordinate system and relevant kinematical variables can be found in Ref.~\cite{Chatrchyan:2008aa}.

The data sample used in this search corresponds to an integrated luminosity
of $\procLumi$ recorded in 2012 with the CMS detector at the
LHC. Dilepton triggers (dielectron, dimuon, muon-electron)
and diphoton triggers are used for data collection.
The transverse momentum (\pt) threshold for
dilepton triggers is 17\GeV for
the leading lepton and 8\GeV for the subleading
lepton. Similarly, the \pt thresholds for the diphoton trigger are 36 and 22\GeV.

The dominant SM backgrounds for this analysis such as $\ttbar$ quark pairs
and diboson production are simulated using the {\MADGRAPH} (version 5.1.3.30)~\cite{Maltoni:2002qb} generator.
We use the CTEQ6L1 leading-order
parton distribution function (PDF) set~\cite{Kretzer:2003it}. For the diboson + jets
simulation, up to two jets are selected at the matrix element level
in {\MADGRAPH}. The detector simulation is performed
with {\GEANTfour}~\cite{Agostinelli:2002hh}. The generation of signal events
is performed using both the {\MADGRAPH} and {\PYTHIA} generators,
with the description of detector response based on the CMS fast simulation
program~\cite{Orbaker:2010zz}.

\section{Particle reconstruction  and preliminary event selection}
\label{objects}

The CMS experiment uses a particle-flow (PF) based event
reconstruction~\cite{CMS-PAS-PFT-09-001,CMS-PAS-PFT-10-001},
which takes into account information from all subdetectors, including
charged-particle tracks from the tracking system and deposited energy from
the electromagnetic and hadronic calorimeters. All
particles in the event are classified into mutually exclusive types:
electrons, muons, $\tau$ leptons, photons, charged hadrons, and neutral hadrons.

Electron and muon candidates used in this search are
reconstructed from the tracker, calorimeter, and muon system
measurements.
Details of reconstruction and identification can be found in
Ref.~\cite{CMS-PAS-EGM-10-004,Chatrchyan:2013dga} for electrons
and in Refs.~\cite{CMS-PAS-MUO-10-002,Chatrchyan:2012xi} for muons.
The electron and muon candidates are required to have $\pt \ge 10\GeV$
and $\abs{\eta} < 2.4$.  For events triggered by the dilepton
trigger, the leading electron or muon must have $\pt > 20\GeV$
in order to ensure maximal efficiency of the dilepton trigger.
Hadronic decays of the  $\tau$ lepton ($\tauh$) are reconstructed
using the hadron-plus-strips (HPS) method~\cite{1748-0221-7-01-P01001}
and must have the measured jet \pt of the jet tagged as a
$\tauh$ candidate to be greater than 20\GeV and $\abs{\eta} \leq 2.3$.

Photon candidates are reconstructed using the energy deposit
clusters in the electromagnetic calorimeter~\cite{CMS-PAS-EGM-10-005, Chatrchyan:2013dga}.
Candidate photons are required to satisfy shower shape
requirements. In order to reject electrons misidentified as photons,
the photon candidate must not match any of the tracks reconstructed
with the pixel detector.  Photon candidates are required to have
$\pt \geq 20\GeV$ and $\abs{\eta} < 2.5$.  For events triggered by the
diphoton trigger, the leading (subleading) photon must have
$\pt > 40 (25)\GeV$.

Jets are reconstructed by clustering PF particles using the
anti-$\kt$ algorithm~\cite{Cacciari:2008gp} with a distance parameter
of 0.5 and are required to have $\abs{\eta} \leq 2.5$.
Jets are further characterized as being ``$\cPqb$-tagged'' using the medium working
point of the CMS Combined secondary-vertex (CSV)
algorithm~\cite{Chatrchyan:2012jua}. They typically result from the decays
of the $\cPqb$ quark.
The total hadronic transverse energy, $\HT$, is the scalar sum of the
$\pt$ of all jets with $\pt > 30\GeV$. The $\MET$ in an event is defined
to be the magnitude of the vectorial \pt sum of all the PF candidates.

The primary vertex for a candidate event is defined as the reconstructed collision vertex with the highest $\pt^{2}$ sum of the associated tracks.
It also must be within 24\unit{cm} from the center of the detector, along the
beam axis ($z$ direction), and within 2\unit{cm} in a direction transverse
to the beam line~\cite{Chatrchyan:2012bra}.
We require the candidate leptons to
originate from within 0.5\unit{cm} in $z$ of the primary vertex and that their
impact parameters $d_{xy}$ between the track and the primary vertex in
the plane transverse to the beam axis be at most 0.02\unit{cm}.

For electrons and muons, we define the
relative isolation $I_{\text{rel}}$ of the candidate leptons to be the ratio
of the \pt sum of all other PF candidates that are reconstructed in a cone defined by
$\Delta{R} = \sqrt{\smash[b]{(\Delta{\eta})^2+(\Delta{\phi})^2}} < 0.3$ around the candidate to
the \pt of the candidate, and require $I_{\text{rel}} < 0.15$.
The photon isolation requirement is similar, but varies as a function of the
candidate \pt and $\eta$~\cite{Khachatryan:2014ira}. For the isolation of the $\tauh$ candidates, we
require that the \pt sum of all other particles in a cone of
$\Delta{R} < 0.5$ be less than 2\GeV. The isolation variable for leptons and
photons is corrected for the contributions from pileup interactions~\cite{Cacciari:2007fd}.
The combined efficiency for trigger, reconstruction and
identification are approximately 75\% for electrons and 80\% for
muons. The identification and isolation efficiency for prompt
leptons is measured in data using a "tag-and-probe" method based
on an inclusive sample of $\cPZ$ + jets events~\cite{CMS:2011aa}.
The ratio of the efficiency in data and simulation parameterized
by the different $\pt$ and $\eta$ values of the probed lepton is
used to correct the selection efficiency in the simulated samples.

A leptonically-decaying $\cPZ$ boson can lead to a trilepton event when the
final-state radiation undergoes (internal or external) conversion and one
of the leptons escapes detection. Therefore, we reject trilepton events with
low missing transverse energy ($\MET < 30\GeV$) when their three body invariant mass
is consistent with the $\cPZ$ mass (i.e. $m_{\ell^{+}\ell^{-}\ell^{\prime\pm}}$ or
$m_{\ell^{+}\ell^{-}\ell^{\pm}}$ is between 75 and 105\GeV), even
if $m_{\ell^+\ell^-}$ is not~($\ell\;= \Pe,\;\Pgm$). Finally, SM background from abundant
low-mass Drell--Yan production and low-mass resonances like \JPsi\
and \PgU\ is suppressed by rejecting an event if it contains  a dilepton pair
with $m_{\ell^+\ell^-}$ below 12\GeV.

\section{Event classification}
\label{strategy}

We perform searches using a multi-channel counting experiment approach.
A multilepton event consists of at least three isolated and prompt
leptons ($\Pe$,\;$\Pgm$,\;$\tauh$), of which at least two must be
electrons or muons (``light'' leptons).
A photon pair together with at least one lepton makes a
lepton+diphoton event. The relatively low rates for multilepton and
lepton+diphoton final states in SM allow this search to target
rare signals.

\subsection{The \texorpdfstring{$\PH \to \Ph \Ph$, $\PA \to \cPZ \Ph$}{H -> h h, A -> Z h}, and \texorpdfstring{$\cPqt \to \cPqc \Ph$}{t -> c h} signals}

In the $\PH \to \Ph \Ph$ search,
seven combinations of the $\Ph \Ph$ decays ($\PW \PW^* \PW \PW^*$,
$\PW \PW^* \cPZ \cPZ^*$, $\PW \PW^* \Pgt \Pgt$, $\cPZ \cPZ^* \cPZ \cPZ^*$,
$\cPZ \cPZ^* \Pgt \Pgt$, $\cPZ \cPZ^* \cPqb \cPqb$, and $\Pgt \Pgt \Pgt \Pgt$)
can result in a final state containing multileptons and three combinations
($\gamma \gamma \PW \PW^*$, $\gamma \gamma \cPZ \cPZ^*$,
and  $\gamma \gamma \Pgt \Pgt$) can result in lepton+diphoton final states
with appreciable rates.

In the $\PA \to \cPZ \Ph$ search, the multilepton and diphoton signal
events can result from the $\PW \PW$, $\cPZ \cPZ$, $\Pgt \Pgt$
and $\gamma \gamma$ decays of the $\Ph$, when accompanied by
the appropriate decays of the $\PW$ and $\cPZ$ bosons and
the $\tau$ lepton. Five combinations of the $\cPZ \Ph$ decays
($\cPZ \to \ell \ell$, $\Ph \to \PW \PW^*$; $\cPZ \to \ell \ell$, $\Ph \to \cPZ \cPZ^*$;
$\cPZ \to \ell \ell$, $\Ph \to \Pgt \Pgt$; $\cPZ \to \nu \nu$, $\Ph \to \cPZ \cPZ^*$;
$\cPZ \to \cPq \cPq$, $\Ph \to \cPZ \cPZ^*$) can result in a final state containing multileptons
and one combination ($\cPZ \to \ell \ell$, $\Ph \to \gamma \gamma$) leads to
lepton+diphoton states with substantial rates.

For the $\cPqt \to \cPqc \Ph$ search, three combinations in the
decay chain $\ttbar \to (\cPqb \PW)(\cPqc \Ph) \to (\cPqb \ell \nu)(\cPqc \Ph)$
can lead to multilepton final states, namely $\Ph \to \PW \PW^*$,
$\Ph \to \cPZ \cPZ^*$, and $\Ph \to \Pgt \Pgt$.
The $\cPqb \PW \cPqc \Ph$ channel can also result in a lepton+diphoton final
state when the Higgs boson decays to a photon pair.
Finally, given the parent $\ttbar$ state, the amount of hadronic activity in
the $\cPqt \to \cPqc \Ph$ signal events is expected to be quite large.

\subsection{Multilepton search channels}

A three-lepton event must contain exactly three
isolated and prompt leptons ($\Pe$,\;\Pgm$,\;\tauh$), of which two
must be electrons or muons. Similarly, a four-lepton event must contain at least four
leptons, of which three must be electrons or muons.
With the goal of segregating SM
backgrounds, these events are classified on the basis
of the lepton flavor, their relative charges, as
well as charge and flavor combinations and other kinematic quantities
such as dilepton invariant mass and $\MET$ , as follows.

Events with $\tauh$ are grouped separately because narrow jets are
frequently misidentified as $\tauh$, leading to larger SM backgrounds for
channels with $\tauh$. Similarly, the presence of a $\cPqb$-tagged jet in an event
calls for a separate grouping in order to isolate the
$\ttbar$ background.

The next classification criterion is the maximum number of opposite-sign
and same-flavor (OSSF) dilepton pairs that can be constructed in an event
using each light lepton only once. For example, both $\Pgmp \Pgmm \Pgmm$
and $\Pgmp \Pgmm \Pem$ are said to be OSSF1, and a $\Pgmp \Pem \tauh$
would be OSSF0. Both $\Pep \Pep \Pgmm$ and $\Pgmp \Pgmp \tauh$
are OSSF0(SS), where SS additionally indicates the presence of same-signed
electron or muon pairs. Similarly, $\Pgmp \Pgmm \Pep \Pem$ is OSSF2.
An event with an OSSF pair is said to be ``on-$\cPZ$'' if the invariant mass of
at least one of the OSSF pair is between 75\GeV and 105\GeV, otherwise it
is ``off-$\cPZ$''. An OSSF1 off-$\cPZ$ event is ``below-$\cPZ$'' or ``above-$\cPZ$''
depending on whether the mass of the pair is less than 75\GeV or more
than 105\GeV, respectively.  An on-$\cPZ$ OSSF2 event may be a
``one on-$\cPZ$'' or a ``two on-$\cPZ$'' event.

Finally, the three-lepton events are classified in five $\MET$ bins: $< 50$,
50--100, 100--150, 150--200, and ${>} 200\GeV$ and the four-lepton
events are classified in four $\MET$ bins: $< 30$, 30--50, 50--100,
and $>$100\GeV. This results in a total of 70 three-lepton channels and 72
four-lepton channels which are listed explicitly when we later present the
tables of event yields and background predictions~(Tables~\ref{tab:results3L} and~\ref{tab:results4L}).

\subsection{Lepton+diphoton search channels}

A diphoton pair together with at least one lepton makes a lepton+diphoton event.
The diphoton invariant mass of the
$\Ph \to \gamma \gamma$
candidates must be between 120 and 130\GeV.
The search channels are
$\gamma \gamma \ell \ell$, $\gamma \gamma \ell \tauh$, $\gamma \gamma \ell$,
and $\gamma \gamma \tauh$. Depending on the relative dilepton flavor and invariant mass,
the $\gamma \gamma \ell \ell$ events can be OSSF0, OSSF1 on-$\cPZ$, or OSSF1 off-$\cPZ$.
The SM background decreases with increasing $\MET$, therefore the events are further classified, when
appropriate, in three bins: $\MET < 30$, 30--50, and $>$50\GeV.

The $\cPqt \to \cPqc \Ph$ signal populates the $\gamma \gamma \ell$
and $\gamma \gamma \tauh$ channels but not the dilepton+diphoton
channels. Since the $\cPqt \to \cPqc \Ph$ signal events always contain
a $\cPqb$ quark from the conventional $\cPqb \PW$ decay of one of the
top quarks, the $\gamma \gamma \ell$ and $\gamma \gamma \tauh$ search channels
are further classified based on the presence of a $\cPqb$-tagged jet.
For these channels, we also split the last $\MET$ bin into two: 50--100$\GeV$
and ${>} 100\GeV$.

The overall lepton+diphoton channel count in this search is seven for $\gamma \gamma \ell \ell$,
three for $\gamma \gamma \ell \tauh$, and eight each for $\gamma \gamma \ell$ and $\gamma \gamma \tauh$.
They are listed explicitly when we present the tables of event yields and background predictions later (Tables~\ref{tab:resultsNobGam2}
and~\ref{tab:resultsYesbGam2}).

\section{Background estimation}
\label{bkgnd}
\subsection{Multilepton background estimation}

Significant sources of multilepton SM background
are $\cPZ$ + jets, diboson production
($VV$ + jets; $V= \PW,\, \cPZ$), $\ttbar$ production, and rare processes
such as $\ttbar V + \text{jets}$.
The techniques we use here to estimate these backgrounds are
identical to those used in Ref.~\cite{Chatrchyan:2014aea}
and are described briefly below.

$\PW \cPZ$ and $\cPZ \cPZ$ diboson production can yield events with three or four
intrinsically prompt and isolated leptons that can be
accompanied by significant $\MET$ and $\HT$.
To estimate these background contributions, we use a simulation validated
after kinematic comparisons with appropriately enriched data samples.

Processes such as $\cPZ$ + jets and $\PWp \PWm$ + jets can
yield events with two prompt leptons. These can be accompanied by jets
that may also contain leptons
from the semileptonic decays of hadrons,
or other objects that are misreconstructed
as prompt leptons, leading to a three-lepton SM background.
Since the simulation of the rare fluctuations that lead to such a
misidentified prompt lepton can be unreliable, we use the data with
two reconstructed leptons to estimate this SM background using the number
of isolated prompt tracks in the dilepton dataset.

The $\ttbar$ decay can result in two prompt leptons and is a source of background
when the decay of one of the daughter $\cPqb$ quarks reconstructs as the third prompt
lepton candidate.
This background is estimated from a $\ttbar$ Monte Carlo sample and using the probability
of occurrence of a misidentified third lepton derived from data.

For search channels that contain $\tauh$, we estimate the probability
of a (sparse) jet misidentified as a $\tauh$ candidate by extrapolating the
isolation distribution of the $\tauh$ candidates.  Since the shape of
this distribution is sensitive to the extent of jet activity, the
extrapolation is carried out as a function of the hadronic activity in
the sample as determined by the summed \pt of all tracks as well
as the leading jet \pt in the event.

Finally, minor backgrounds from rare processes such
as $\ttbar V + \text{jets}$ or SM Higgs production including its associated
production with $\PW$, $\cPZ$, or $\ttbar$ are estimated using simulation.

\subsection{Lepton+diphoton background estimation}

We use a 120--130\GeV diphoton invariant mass window to capture
the $\Ph \to \gamma \gamma $ signal. With the requirement of at
least one lepton in these lepton+diphoton channels,
the SM background tends to be small and is estimated by interpolating
the diphoton mass sidebands of the signal window. We assume the
background distribution shape to be a falling exponential as a
function of the diphoton invariant mass over the 100--200\GeV
mass range.

\begin{figure}[!ht]
\centering
\includegraphics[width=0.46 \textwidth]{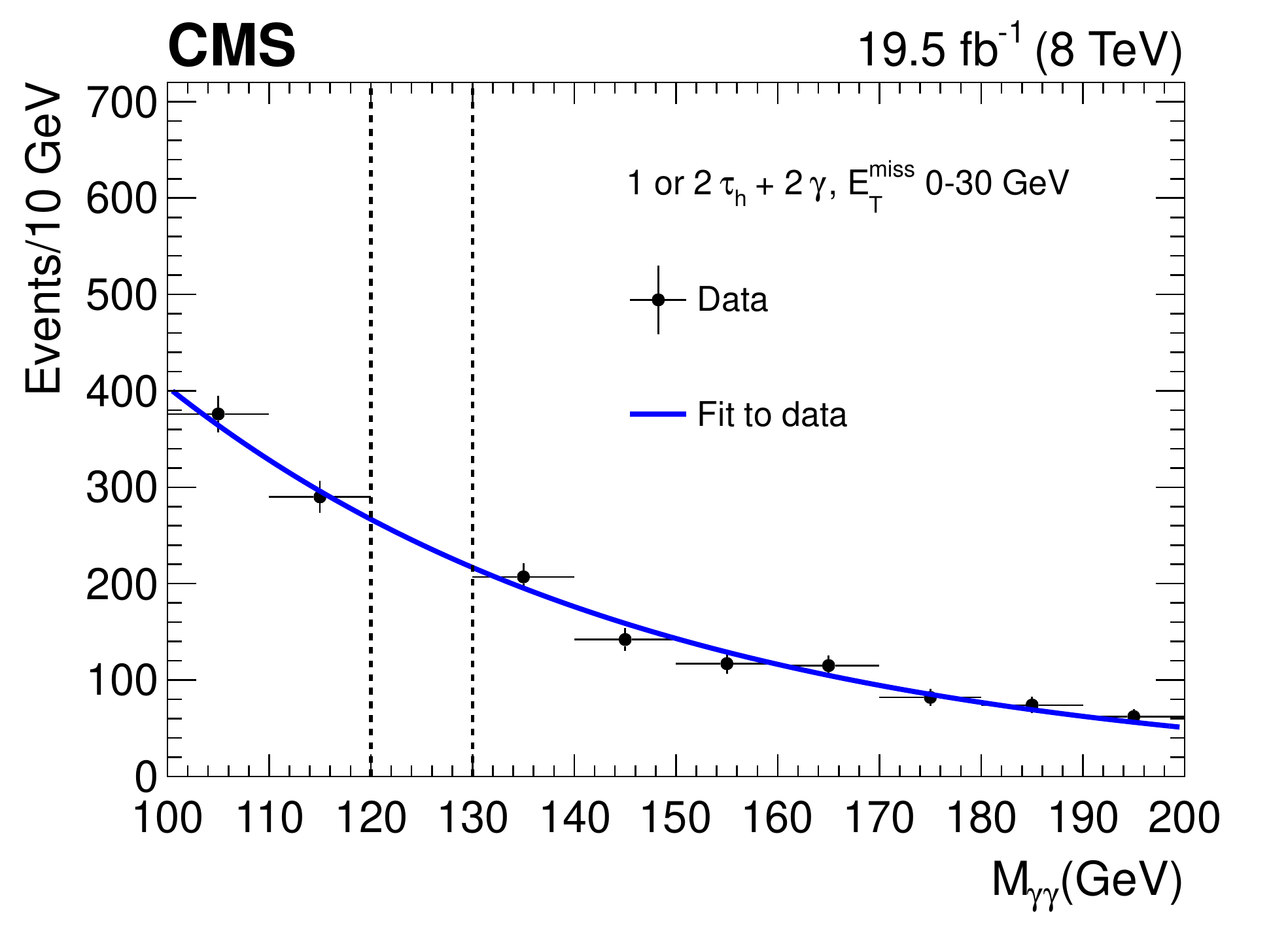}
\includegraphics[width=0.46 \textwidth]{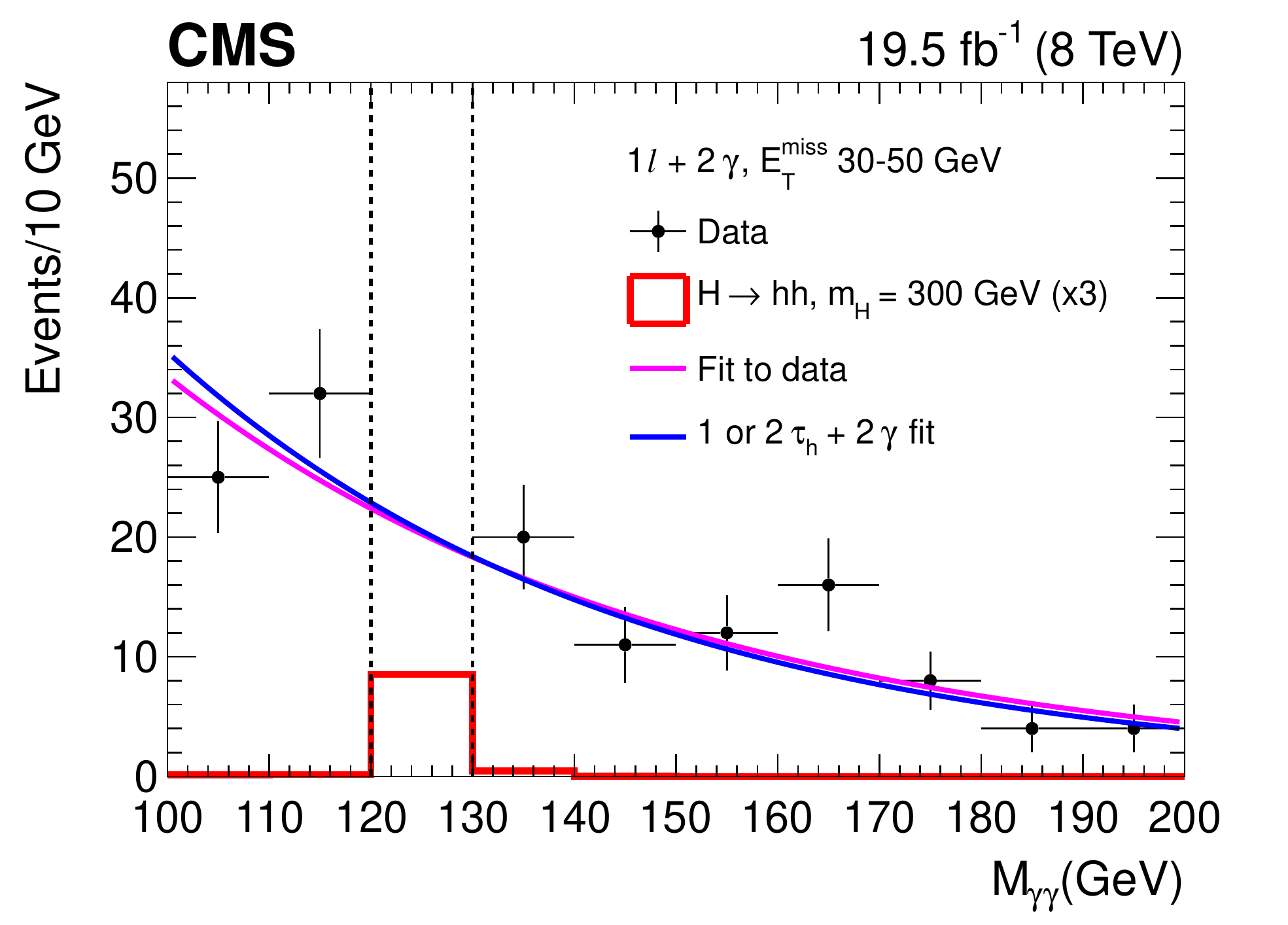}
\caption{\cmsLLeft: diphoton invariant mass distribution for $\gamma \gamma \tauh$
events with $\MET < 30\GeV$ with an exponential fit derived from the 100--120 and 130--200\GeV
sidebands regions. \cmsRRight: the same distribution for the $\gamma \gamma \ell$ events with $\MET$ in 30--50\GeV
range with an exponential fit (blue curve) where the exponent is fixed to the value obtained from the fit
shown in the \cmsLeft~figure.  Also shown for comparison purposes is an actual fit (magenta curve) to the shown
data distribution. An example signal distribution (in red),
assuming $\sigma \mathcal{B}(\Pp \Pp \to \PH \to \Ph \Ph)$ to be equal to three times 3.59\unit{pb},
as described in the text, shows that
the signal is well-contained in the 120--130\GeV window.
}
\label{fig:L1Gam2Fits}
\end{figure}

Figure~\ref{fig:L1Gam2Fits}~(\cmsLLeft) shows the exponential fit
to the 100--120 and 130--200\GeV sidebands in the mass distribution
for $\gamma \gamma \tauh$ events with $\MET < 30\GeV$. We choose this
sample to determine the exponent because it is a high-statistics sample.
This exponent is used for background determination in all diphoton channels,
allowing only the normalization to float from channel to channel. Figure~\ref{fig:L1Gam2Fits}~(\cmsRRight)
shows an example of such a fit for the $\gamma \gamma \ell$ sample
with a 30--50\GeV $\MET$ requirement along with an exponential
fit where both the exponent and normalization are allowed to float.
We assign a 50\% systematic uncertainty for background determination in the 120--130\GeV Higgs boson
mass region. The figure also shows
the expected signal multiplied by a factor of three for clarity
for $m_{\PH} = 300\GeV$, assuming that the
production cross section $\sigma$ for $m_{\PH} = 300\GeV$ is
equal to the Standard Model Higgs
gluon fusion value of 3.59\unit{pb} at this mass given by the LHC Higgs
Cross Section Working Group in Ref.~\cite{Heinemeyer:2013tqa}, and a
branching fraction $\mathcal{B}(\PH \to \Ph \Ph)=1$.

\section{Observations}
\label{obs}

Tables~\ref{tab:results3L} and~\ref{tab:results4L} list the observed
number of events for the three-lepton and four-lepton search channels,
respectively. The number of expected events from SM processes are also shown together
with the combined statistical and systematic uncertainties.
Table~\ref{tab:systematics} lists the sources of systematic effects and
the resultant uncertainties in estimating the expected events from the SM.
All search channels share systematic
uncertainties for luminosity, renormalization scale, PDF, and trigger
efficiency.

\begin{table*}[!htp]
\centering
\topcaption{Observed (Obs.) yields and SM expectations (Exp.) for three-lepton events. See text for the
description of event classification by the number and invariant mass of opposite-sign, same-flavor
lepton pairs that are on- or below-$\cPZ$ (see Section 4.2), presence of $\tauh$, tagged $\cPqb$ jets,
and the $\MET$ in the event. The 70 channels are exclusive.}
\label{tab:results3L}
\resizebox{\textwidth}{!}{
\begin{scotch}{clc|cc|cc|cc|cc}
\multicolumn{1}{c}{3 leptons} & $m_{\ell^+\ell^-}$ & $\MET$ & \multicolumn{2}{c|}{${N}_{\tauh} = 0$, ${N}_{\cPqb} = 0$} & \multicolumn{2}{c|}{${N}_{\tauh} = 1$, ${N}_{\cPqb} = 0$} & \multicolumn{2}{c|}{${N}_{\tauh} = 0$, ${N}_{\cPqb} \geq 1$} & \multicolumn{2}{c}{${N}_{\tauh} = 1$, ${N}_{\cPqb} \geq 1$} \\
& & (\GeVns) & Obs. & Exp. & Obs. & Exp. & Obs. & Exp. & Obs. & Exp. \\
\hline
OSSF0(SS) & \NA & (200, $\infty$) & 1 & 1.3 $\pm$ 0.6 & 2 & 1.4 $\pm$ 0.5 & 0 & 0.70 $\pm$ 0.36 & 0 & 0.7 $\pm$ 0.5 \\
OSSF0(SS) & \NA & (150, 200) & 2 & 2.1 $\pm$ 0.9 & 0 & 3.0 $\pm$ 1.1 & 1 & 2.1 $\pm$ 1.0 & 0 & 1.5 $\pm$ 0.6 \\
OSSF0(SS) & \NA & (100, 150) & 9 & 10.0 $\pm$ 4.9 & 4 & 9.9 $\pm$ 3.0 & 12 & 12.0 $\pm$ 5.9 & 4 & 6.3 $\pm$ 2.8 \\
OSSF0(SS) & \NA & (50, 100) & 34 & 37 $\pm$ 15 & 54 & 66 $\pm$ 14 & 32 & 32 $\pm$ 15 & 24 & 22 $\pm$ 10 \\
OSSF0(SS) & \NA & (0, 50) & 47 & 46 $\pm$ 11 & 196 & 221 $\pm$ 51 & 28 & 24 $\pm$ 11 & 21 & 31.0 $\pm$ 9.6 \\
OSSF0 & \NA & (200, $\infty$) & \NA & \NA & 5 & 4.8 $\pm$ 2.4 & \NA & \NA & 6 & 5.9 $\pm$ 3.1 \\
OSSF0 & \NA & (150, 200) & \NA & \NA & 12 & 18.0 $\pm$ 9.1 & \NA & \NA & 21 & 20 $\pm$ 10 \\
OSSF0 & \NA & (100, 150) & \NA & \NA & 94 & 96 $\pm$ 47 & \NA & \NA & 91 & 121 $\pm$ 61 \\
OSSF0 & \NA & (50, 100) & \NA & \NA & 351 & 329 $\pm$ 173 & \NA & \NA & 300 & 322 $\pm$ 163 \\
OSSF0 & \NA & (0, 50) & \NA & \NA & 682 & 767 $\pm$ 207 & \NA & \NA & 230 & 232 $\pm$ 118 \\
OSSF1 & Below-$\cPZ$ & (200, $\infty$) & 2 & 2.5 $\pm$ 0.9 & 4 & 2.1 $\pm$ 1.0 & 1 & 1.9 $\pm$ 0.7 & 2 & 2.4 $\pm$ 1.2 \\
OSSF1 & On-$\cPZ$  & (200, $\infty$) & 17 & 19.0 $\pm$ 6.3 & 4 & 5.6 $\pm$ 1.9 & 1 & 2.4 $\pm$ 0.8 & 3 & 2.1 $\pm$ 0.9 \\
OSSF1 & Below-$\cPZ$  & (150, 200) & 7 & 4.4 $\pm$ 1.7 & 11 & 9.3 $\pm$ 4.6 & 3 & 4.7 $\pm$ 2.1 & 7 & 11.0 $\pm$ 5.9 \\
OSSF1 & On-$\cPZ$ & (150, 200) & 38 & 32.0 $\pm$ 8.5 & 10 & 11.0 $\pm$ 3.6 & 4 & 5.4 $\pm$ 1.7 & 2 & 5.7 $\pm$ 2.7 \\
OSSF1 & Below-$\cPZ$  & (100, 150) & 21 & 26.0 $\pm$ 9.9 & 45 & 56 $\pm$ 27 & 20 & 23 $\pm$ 11 & 56 & 66 $\pm$ 33 \\
OSSF1 & On-$\cPZ$ & (100, 150) & 134 & 129 $\pm$ 29 & 43 & 51 $\pm$ 16 & 20 & 18 $\pm$ 6 & 24 & 28 $\pm$ 14 \\
OSSF1 & Below-$\cPZ$  & (50, 100) & 157 & 129 $\pm$ 30 & 383 & 380 $\pm$ 104 & 58 & 60 $\pm$ 28 & 166 & 173 $\pm$ 87 \\
OSSF1 & On-$\cPZ$ & (50, 100) & 862 & 732 $\pm$ 141 & 1360 & 1230 $\pm$ 323 & 80 & 62 $\pm$ 17 & 117 & 101 $\pm$ 48 \\
OSSF1 & Below-$\cPZ$  & (0, 50) & 543 & 559 $\pm$ 93 & 10200 & 9170 $\pm$ 2710 & 40 & 52 $\pm$ 14 & 257 & 256 $\pm$ 79 \\
OSSF1 & On-$\cPZ$ & (0, 50) & 4040 & 4060 $\pm$ 691 & 51400 & 51400 $\pm$ 15300 & 181 & 181 $\pm$ 28 & 1000 & 1010 $\pm$ 286 \\
\end{scotch}
}
\end{table*}

\begin{table*}[!htbp]
\centering
\topcaption{Observed (Obs.) yields and SM expectation (Exp.) for four-lepton events. See text for
the description of event classification by the number and invariant mass of opposite-sign, same-flavor
lepton pairs that are on- or off-$\cPZ$, presence of $\tauh$, tagged $\cPqb$ jets, and the total $\MET$
in the event. The 72 channels are exclusive.}
\label{tab:results4L}
\resizebox{\textwidth}{!}{
\begin{scotch}{clc|cc|cc|cc|cc}
\multicolumn{1}{c}{$\geq$4 leptons} & $m_{\ell^+\ell^-}$ & $\MET$ & \multicolumn{2}{c|}{${N}_{\tauh} = 0$, ${N}_{\cPqb} = 0$} & \multicolumn{2}{c|}{${N}_{\tauh} = 1$, ${N}_{\cPqb} = 0$} & \multicolumn{2}{c|}{${N}_{\tauh} = 0$, ${N}_{\cPqb} \geq 1$} & \multicolumn{2}{c}{${N}_{\tauh} = 1$, ${N}_{\cPqb} \geq 1$} \\
& & (\GeVns) & Obs. & Exp. & Obs. & Exp. & Obs. & Exp. & Obs. & Exp. \\
\hline
OSSF0 & \NA & (100, $\infty$) & 0 & 0.07 $\pm$ 0.07 & 0 & 0.18 $\pm$ 0.09 & 0 & 0.05 $\pm$ 0.05 & 0 & 0.16 $\pm$ 0.10 \\
OSSF0 & \NA & (50, 100) & 0 & 0.07 $\pm$ 0.06 & 2 & 0.80 $\pm$ 0.35 & 0 & $0.00^{+0.03}_{-0.00}$ & 0 & 0.43 $\pm$ 0.22 \\
OSSF0 & \NA & (30, 50) & 0 & $0.001^{+0.020}_{-0.001}$ & 0 & 0.47 $\pm$ 0.24 & 0 & $0.00^{+0.02}_{-0.00}$ & 0 & 0.11 $\pm$ 0.09 \\
OSSF0 & \NA & (0, 30) & 0 & $0.007^{+0.020}_{-0.007}$ & 1 & 0.40 $\pm$ 0.16 & 0 & $0.001^{+0.020}_{-0.001}$ & 0 & $0.02^{+0.04}_{-0.02}$ \\
OSSF1 & Off-$\cPZ$ & (100, $\infty$) & 0 & 0.07 $\pm$ 0.04 & 4 & 1.00 $\pm$ 0.33 & 0 & 0.14 $\pm$ 0.09 & 0 & 0.46 $\pm$ 0.20 \\
OSSF1 & On-$\cPZ$ & (100, $\infty$) & 2 & 0.6 $\pm$ 0.2 & 2 & 3.4 $\pm$ 0.8 & 1 & 0.80 $\pm$ 0.41 & 0 & 0.60 $\pm$ 0.26 \\
OSSF1 & Off-$\cPZ$ & (50, 100) & 0 & 0.21 $\pm$ 0.09 & 5 & 2.6 $\pm$ 0.6 & 0 & 0.21 $\pm$ 0.11 & 1 & 0.70 $\pm$ 0.32 \\
OSSF1 & On-$\cPZ$ & (50, 100) & 2 & 1.30 $\pm$ 0.39 & 10 & 12.0 $\pm$ 2.5 & 2 & 0.60 $\pm$ 0.33 & 1 & 0.8 $\pm$ 0.3 \\
OSSF1 & Off-$\cPZ$ & (30, 50) & 1 & 0.16 $\pm$ 0.07 & 4 & 2.4 $\pm$ 0.5 & 0 & 0.06 $\pm$ 0.06 & 0 & 0.47 $\pm$ 0.21 \\
OSSF1 & On-$\cPZ$ & (30, 50) & 3 & 1.20 $\pm$ 0.35 & 11 & 14.0 $\pm$ 3.1 & 0 & 0.22 $\pm$ 0.12 & 0 & 0.80 $\pm$ 0.31 \\
OSSF1 & Off-$\cPZ$ & (0, 30) & 1 & 0.38 $\pm$ 0.18 & 11 & 5.7 $\pm$ 1.7 & 0 & 0.05 $\pm$ 0.04 & 0 & 0.50 $\pm$ 0.26 \\
OSSF1 & On-$\cPZ$ & (0, 30) & 1 & 2.0 $\pm$ 0.5 & 32 & 30.0 $\pm$ 9.2 & 1 & 0.19 $\pm$ 0.11 & 3 & 1.30 $\pm$ 0.42 \\
OSSF2 & Two on-$\cPZ$ & (100, $\infty$) & 0 & 0.02 $\pm$ 0.15 & \NA & \NA & 0 & 0.21 $\pm$ 0.13 & \NA & \NA \\
OSSF2 & One on-$\cPZ$ & (100, $\infty$) & 1 & 0.43 $\pm$ 0.15 & \NA & \NA & 0 & 0.50 $\pm$ 0.29 & \NA & \NA \\
OSSF2 & Off-$\cPZ$ & (100, $\infty$) & 0 & 0.06 $\pm$ 0.03 & \NA & \NA & 0 & 0.09 $\pm$ 0.07 & \NA & \NA \\
OSSF2 & Two on-$\cPZ$ & (50, 100) & 3 & 2.8 $\pm$ 2.1 & \NA & \NA & 0 & 0.33 $\pm$ 0.11 & \NA & \NA \\
OSSF2 & One on-$\cPZ$ & (50, 100) & 1 & 2.0 $\pm$ 0.7 & \NA & \NA & 1 & 0.50 $\pm$ 0.28 & \NA & \NA \\
OSSF2 & Off-$\cPZ$ & (50, 100) & 2 & 0.20 $\pm$ 0.14 & \NA & \NA & 0 & 0.12 $\pm$ 0.10 & \NA & \NA \\
OSSF2 & Two on-$\cPZ$ & (30, 50) & 19 & 22 $\pm$ 9 & \NA & \NA & 2 & 0.70 $\pm$ 0.24 & \NA & \NA \\
OSSF2 & One on-$\cPZ$ & (30, 50) & 6 & 6.5 $\pm$ 2.4 & \NA & \NA & 0 & 0.32 $\pm$ 0.12 & \NA & \NA \\
OSSF2 & Off-$\cPZ$ & (30, 50) & 3 & 1.4 $\pm$ 0.6 & \NA & \NA & 1 & 0.15 $\pm$ 0.08 & \NA & \NA \\
OSSF2 & Two on-$\cPZ$ & (0, 30) & 118 & 109 $\pm$ 28 & \NA & \NA & 3 & 2.0 $\pm$ 0.5 & \NA & \NA \\
OSSF2 & One on-$\cPZ$ & (0, 30) & 24 & 29.0 $\pm$ 7.6 & \NA & \NA & 1 & 0.60 $\pm$ 0.17 & \NA & \NA \\
OSSF2 & Off-$\cPZ$ & (0, 30) & 5 & 7.8 $\pm$ 2.3 & \NA & \NA & 0 & 0.18 $\pm$ 0.06 & \NA & \NA \\
\end{scotch}
}
\end{table*}

\begin{table}
\centering
\topcaption{A compilation of significant sources of systematic uncertainties
in the event yield estimation. Note that a given uncertainty may pertain
only to specific sources of background. The listed values are representative
and the impact of an uncertainty
varies from search channel to channel.}
\label{tab:systematics}
\begin{scotch}{lc}
Source of uncertainty & Magnitude (\%) \\
\hline
Luminosity & 2.6 \\
PDF & 10 \\
$\MET (> 50 \GeV)$ resolution correction & 4 \\
Jet energy scale & 0.5 \\
$\cPqb$-tag scale factor ($\ttbar$) & 6 \\
$\Pe (\Pgm$) ID/isolation (at $\pt = 30 \GeV$) & 0.6 (0.2)\\
Trigger efficiency & 5 \\
$\ttbar$ misidentification & 50 \\
$\ttbar, \PW \cPZ, \cPZ \cPZ$ cross sections & 10, 15, 15 \\
$\tauh$ misidentification & 30 \\
Diphoton background & 50 \\
\end{scotch}
\end{table}

The observations listed in the tables generally agree with the
expectations within the uncertainties. Given the
large number of channels being investigated
simultaneously, certain deviations between observations and expected
values are to be anticipated. We discuss one such deviation later in the
context of the $\PH$ search.

\begin{figure}[!ht]
\centering
\includegraphics[width=0.46 \textwidth]{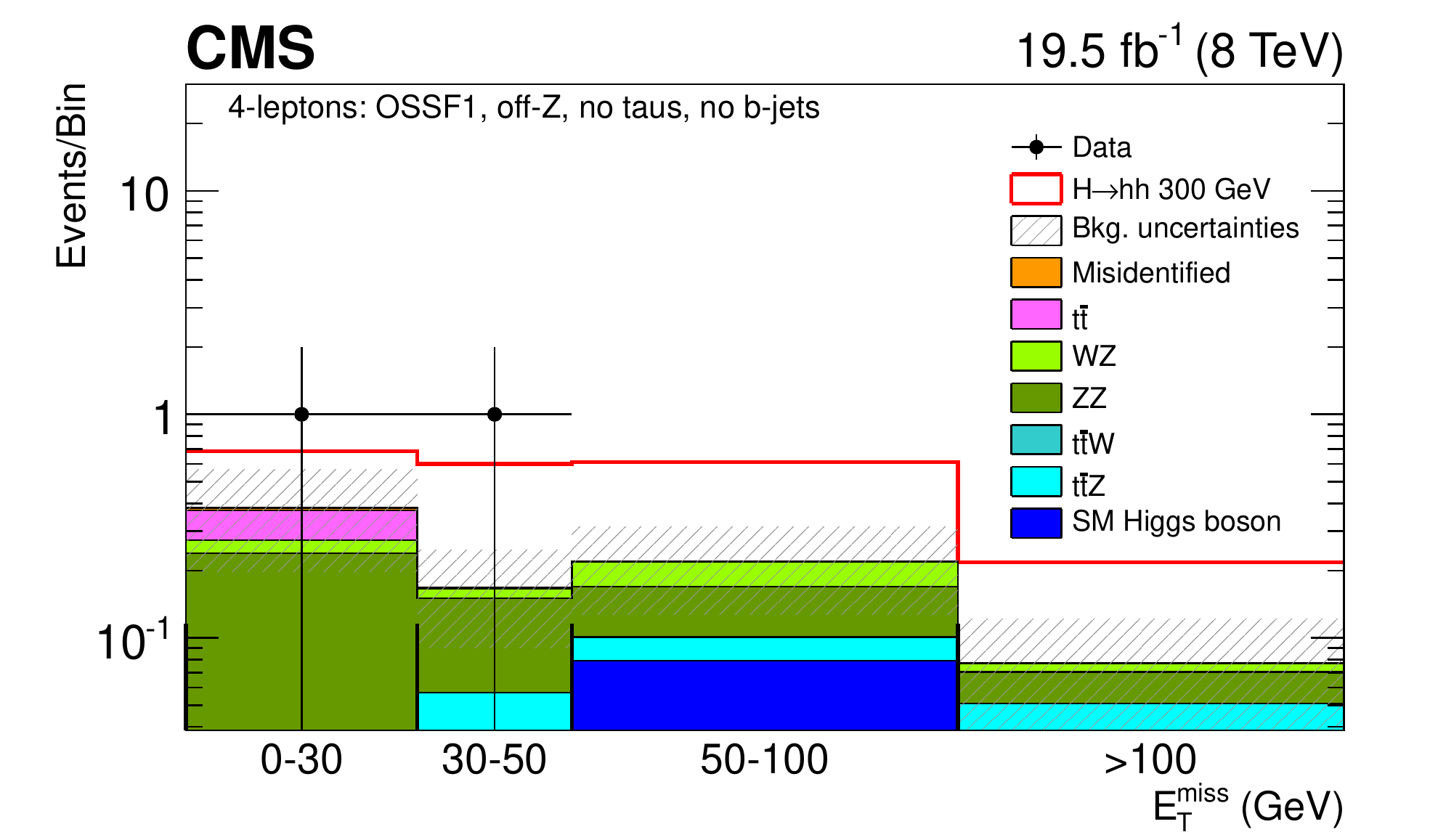}
\includegraphics[width=0.46 \textwidth]{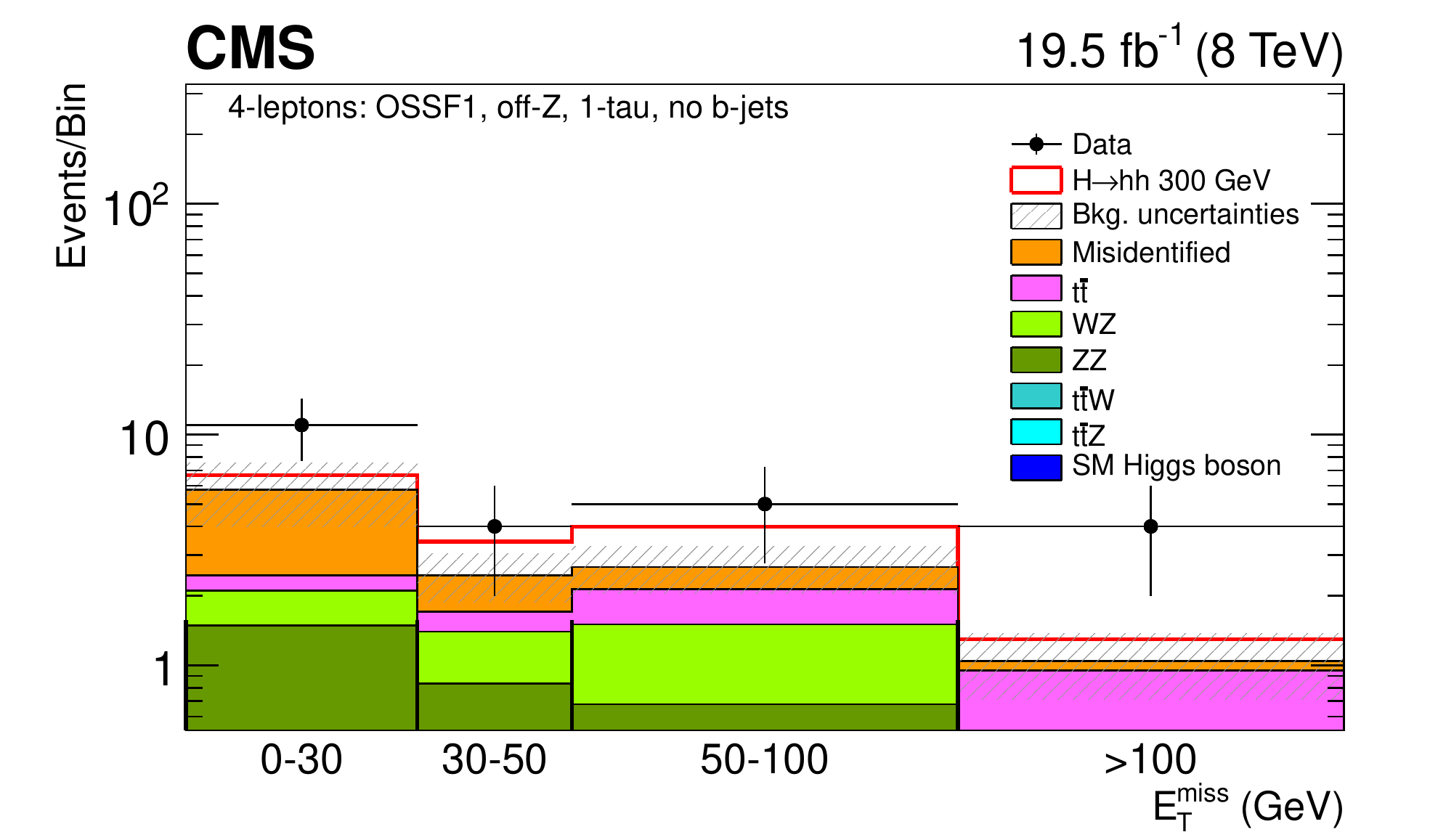}
\caption{The $\MET$ distributions for four-lepton events with an off-$\cPZ$ OSSF1 dilepton pair, no $\cPqb$-tagged jet,
and no $\tauh$ (\cmsLeft), and one $\tauh$~(\cmsRight). These non-resonant (off-$\cPZ$) channels are sensitive to
the $\PH \to \Ph \Ph$ signal which is shown stacked on top of the background distributions, assuming
$\sigma \mathcal{B}(\Pp \Pp \to \PH \to \Ph \Ph) = 3.59$\unit{pb}, as described in the text.}
\label{fig:ResultsHhh}
\end{figure}

\begin{figure}[!ht]
\centering\includegraphics[width=0.46 \textwidth]{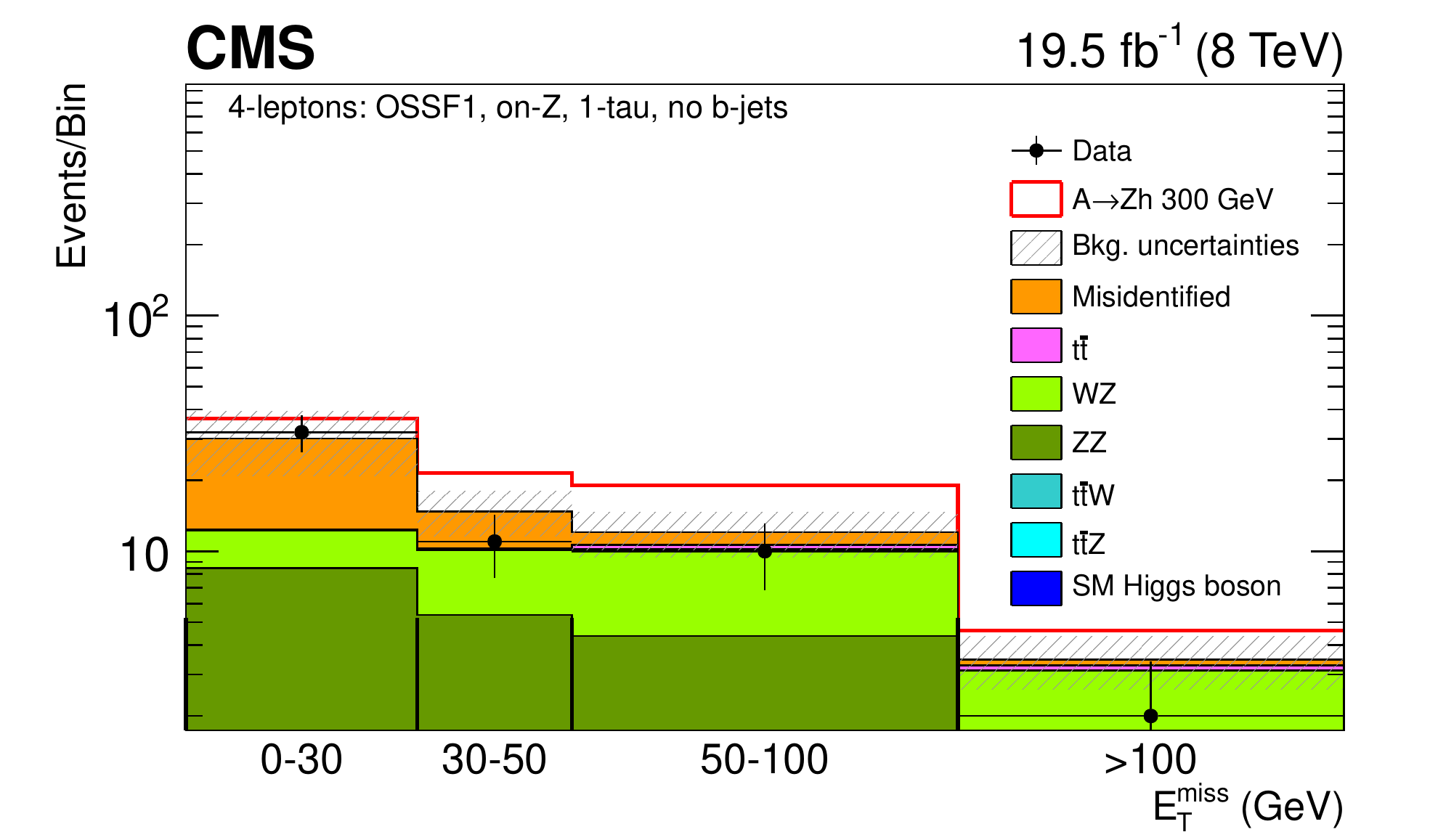}
\includegraphics[width=0.46 \textwidth]{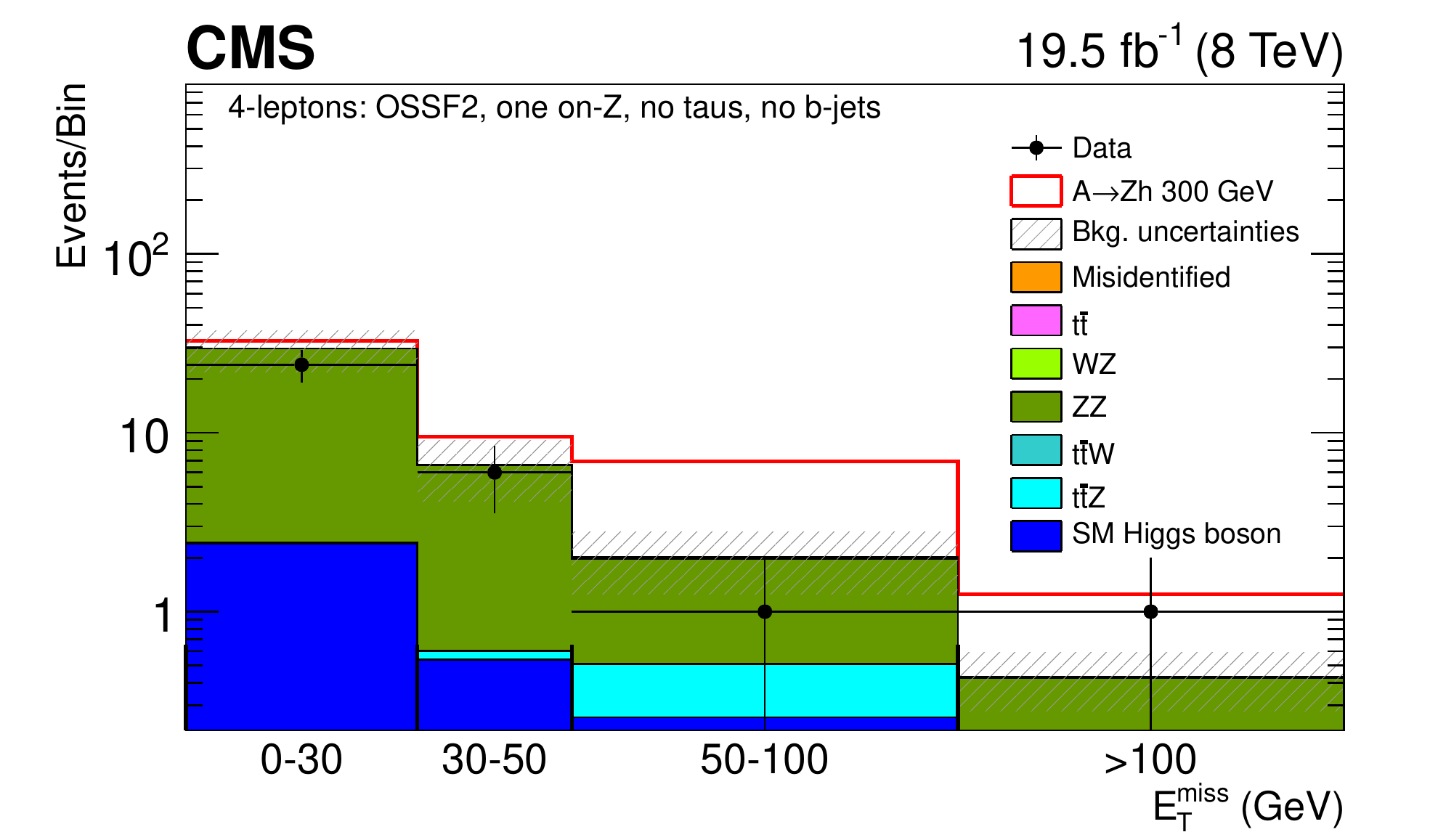}
\caption{The $\MET$ distributions for four-lepton events without $\cPqb$-tagged jets which contain an
on-$\cPZ$ OSSF1 dilepton pair and one $\tauh$ (\cmsLeft), and an OSSF2 dilepton pairs with one $\cPZ$
candidate and no $\tauh$~(\cmsRight). These resonant (containing a $\cPZ$) channels are sensitive to the
$\PA \to \cPZ \Ph$ signal which is shown stacked on top of the background distributions,
assuming $\sigma \mathcal{B}(\Pp \Pp \to \PA \to \cPZ \Ph) = 3.59$\unit{pb}, as described in the text.}
\label{fig:ResultsAZh}
\end{figure}

Figure~\ref{fig:ResultsHhh} shows observations and background decomposition
for some of the most sensitive channels for the $\PH \to \Ph \Ph$ search.
The amount of signal for $m_{\PH} = 300\GeV$, as described above in the context
of Fig.~\ref{fig:L1Gam2Fits}, is also shown. This information is also listed in Table~\ref{tab:resultsHeavyHiggs}.
Figure~\ref{fig:ResultsAZh} and Table~\ref{tab:resultsAtoZHiggs} shows the same for the $\PA \to \cPZ \Ph$
search for $m_{\PA} = 300\GeV$, assuming the same cross section and
$\mathcal{B}(\PA \to \cPZ \Ph)=1$.

\begin{table}[!htb]
\centering
\topcaption{Observed (Obs.) yields and SM expectation (Exp.) for selected four-lepton
channels in the $\PH \to \Ph \Ph$ search. These are also shown in Fig.~\ref{fig:ResultsHhh}.
See text for the description of event classification. The $\PH \to \Ph \Ph$ signal (Sig.) is also
listed, assuming $\sigma \mathcal{B}(\Pp \Pp \to \PH \to \Ph \Ph) = 3.59$\unit{pb}.}
\label{tab:resultsHeavyHiggs}
\begin{scotch}{c|c|ccc}
Channel & $\MET$ (\GeVns) & Obs. & Exp. & Sig.\\
\hline
\multirow{4}{*}{$\genfrac{}{}{0pt}{0}{4\ell~\text{(OSSF1, off-$\cPZ$)}}{\text{(no}~\tauh \text{,~no b-jets)}}$} & (0, 30) & 1 & 0.38 $\pm$ 0.18 & 0.30 \\
& (30, 50) & 1 & 0.16 $\pm$ 0.07 & 0.43 \\
& (50, 100) & 0 & 0.21 $\pm$ 0.09 & 0.39 \\
& (100, $\infty$) & 0 & 0.07 $\pm$ 0.04 & 0.14 \\
\hline
\multirow{4}{*}{$\genfrac{}{}{0pt}{0}{4\ell~\text{(OSSF1, off-$\cPZ$)}}{\text{(1-}~\tauh \text{,~no b-jets)}}$} & (0, 30) & 11 & 5.7 $\pm$ 1.7 & 0.91 \\
& (30, 50) & 4 & 2.4 $\pm$ 0.5 & 0.98 \\
& (50, 100) & 5 & 2.6 $\pm$ 0.6 & 1.31 \\
& (100, $\infty$) & 4 & 1.00 $\pm$ 0.33 & 0.25 \\
\end{scotch}
\end{table}

\begin{table}[!htb]
\centering
\topcaption{Observed (Obs.) yields and SM expectation (Exp.) for selected four-lepton
channels in the $\PA \to \cPZ \Ph$ search. These are also shown in Fig.~\ref{fig:ResultsAZh}.
See text for the description of event classification. The $\PA \to \cPZ \Ph$ signal (Sig.) is also
listed, assuming $\sigma \mathcal{B}(\Pp \Pp \to \PA \to \cPZ \Ph) = 3.59$\unit{pb}.}
\label{tab:resultsAtoZHiggs}
\begin{scotch}{c|c|ccc}
Channel & $\MET$ (\GeVns) & Obs. & Exp. & Sig.\\
\hline
\multirow{4}{*}{$\genfrac{}{}{0pt}{0}{4\ell~\text{(OSSF1, on-$\cPZ$)}}{\text{(1-}~\tauh \text{,~no b-jets)}}$} & (0, 30) & 32 & 30.0 $\pm$ 9.2 & 6.46 \\
& (30, 50) & 11 & 14.0 $\pm$ 3.1 & 6.72 \\
& (50, 100) & 10 & 12.0 $\pm$ 2.5 & 7.05 \\
& (100, $\infty$) & 2 & 3.4 $\pm$ 0.8 & 1.12 \\
\hline
\multirow{4}{*}{$\genfrac{}{}{0pt}{0}{4\ell~\text{(OSSF2, one on-$\cPZ$)}}{(\text{no}~\tauh, \text{~no b-jets)}}$} & (0, 30) & 24 & 29.0 $\pm$ 7.6 & 3.15 \\
& (30, 50) & 6 & 6.5 $\pm$ 2.4 & 2.91 \\
& (50, 100) & 1 & 2.0 $\pm$ 0.7 & 4.92 \\
& (100, $\infty$) & 1 & 0.43 $\pm$ 0.15 & 0.82 \\
\end{scotch}
\end{table}

The lepton+diphoton results are summarized in
Tables~\ref{tab:resultsNobGam2} and~\ref{tab:resultsYesbGam2}. The
observations agree with the expectations
within the uncertainties.

\begin{table}[!htb]
\centering
\topcaption{Observed yields and SM expectations for dilepton+diphoton events.
The diphoton invariant mass is required to be in the 120--130\GeV window.
The ten channels are exclusive.}
\label{tab:resultsNobGam2}
\begin{scotch}{c|c|cc}
Channel & $\MET$ (\GeVns) & Obs. & Exp. \\
\hline
\multirow{3}{*}{$\genfrac{}{}{0pt}{0}{\gamma \gamma \ell \ell}{\text{(OSSF1, off-$\cPZ$)}}$} & (50, $\infty$) & 0 & $0.19^{+0.25}_{-0.19}$ \\
& (30, 50) & 1& $0.17^{+0.25}_{-0.17}$ \\
& (0, 30)  & 1& 1.20 $\pm$ 0.74 \\
\hline
\multirow{3}{*}{$\genfrac{}{}{0pt}{0}{\gamma \gamma \ell \ell}{\text{(OSSF1, on-$\cPZ$)}}$} & (50, $\infty$) & 0 & $0.10^{+0.17}_{-0.10}$ \\
& (30, 50) & 1 & 0.33 $\pm$ 0.28 \\
& (0, 30) & 0 & 1.01 $\pm$ 0.55 \\
\hline
$\genfrac{}{}{0pt}{0}{\gamma \gamma \ell \ell}{\text{(OSSF0)}}$ & All & 0 & $0.00^{+0.17}_{-0.00}$ \\
\hline
\multirow{3}{*}{$\gamma \gamma \ell \tauh$} & (50, $\infty$) & 0 & $0.16^{+0.66}_{-0.16}$ \\
& (30, 50) & 0 & $0.50^{+0.57}_{-0.50}$ \\
& (0, 30) & 0 & 0.76 $\pm$ 0.60 \\
\end{scotch}
\end{table}

\begin{table}[!htb]
\centering
\topcaption{Observed yields and SM expectations for single lepton+diphoton events.
The diphoton invariant mass is required to be in the 120--130\GeV window.
The eight channels are exclusive.}
\label{tab:resultsYesbGam2}
\begin{scotch}{c|c|cc|cc}
Channel & $\MET$ (\GeVns) & \multicolumn{2}{c|}{${N}_{\cPqb} = 0$} & \multicolumn{2}{c}{${N}_{\cPqb} \ge 1$} \\
& & Obs. & Exp. & Obs. & Exp. \\
\hline
\multirow{4}{*}{$\gamma \gamma \ell$} & (100, $\infty$) & 1 & 2.2 $\pm$ 1.0 & 0 & 0.5 $\pm$ 0.4 \\
& (50, 100) & 7 & 9.5 $\pm$ 4.4 &1 & 2.3 $\pm$ 1.2 \\
& (30, 50) & 29 & 21 $\pm$ 10 & 2 & 1.1 $\pm$ 0.6 \\
& (0, 30) & 72 & 77 $\pm$ 38 & 2 & 2.1 $\pm$ 1.1 \\
\hline
\multirow{4}{*}{$\gamma \gamma \tauh$} & (100, $\infty$) & 1 & $0.24^{+0.25}_{-0.24}$ & 0 & 0.35 $\pm$ 0.28 \\
& (50, 100) & 14 & 9.3 $\pm$ 4.7 & 1 & 1.5 $\pm$ 0.8 \\
& (30, 50) & 71 & 67 $\pm$ 34 & 2 & 2.1 $\pm$ 1.2 \\
& (0, 30) & 229 & 235 $\pm$ 117 & 6 & 6.4 $\pm$ 3.3 \\
\end{scotch}
\end{table}

\section{Interpretation of results}
\label{interpr}
\subsection{Statistical procedure}

No significant disagreement is found between our observations and the corresponding SM expectations.
We derive limits on the production cross-section times branching fraction
for the new physics scenarios under consideration, and use them to constrain parameters
of the models.
We set 95\% \CL upper limits on the
cross sections using the modified frequentist construction \CL~\cite{Junk:1999kv,Read:2002hq}. 
We compute the single-channel \CL limits for each channel
and then obtain the combined upper limit.

\subsection{\texorpdfstring{$\PH \to \Ph \Ph$}{H -> h h} and \texorpdfstring{$\PA \to \cPZ \Ph$}{A -> c h} model-independent interpretations}

\begin{figure}[!ht]
\centering
\includegraphics[width=0.46 \textwidth]{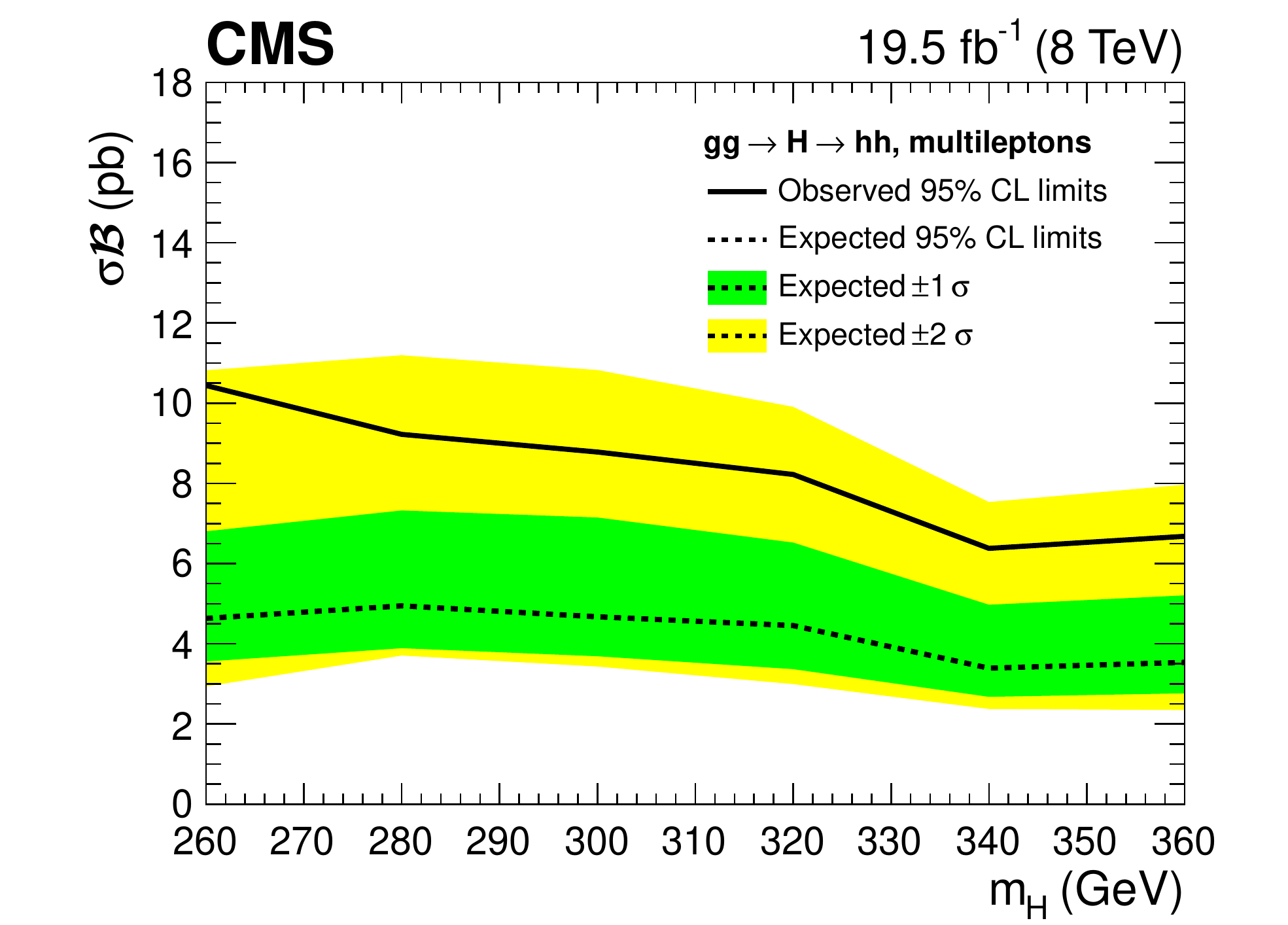}
\includegraphics[width=0.46 \textwidth]{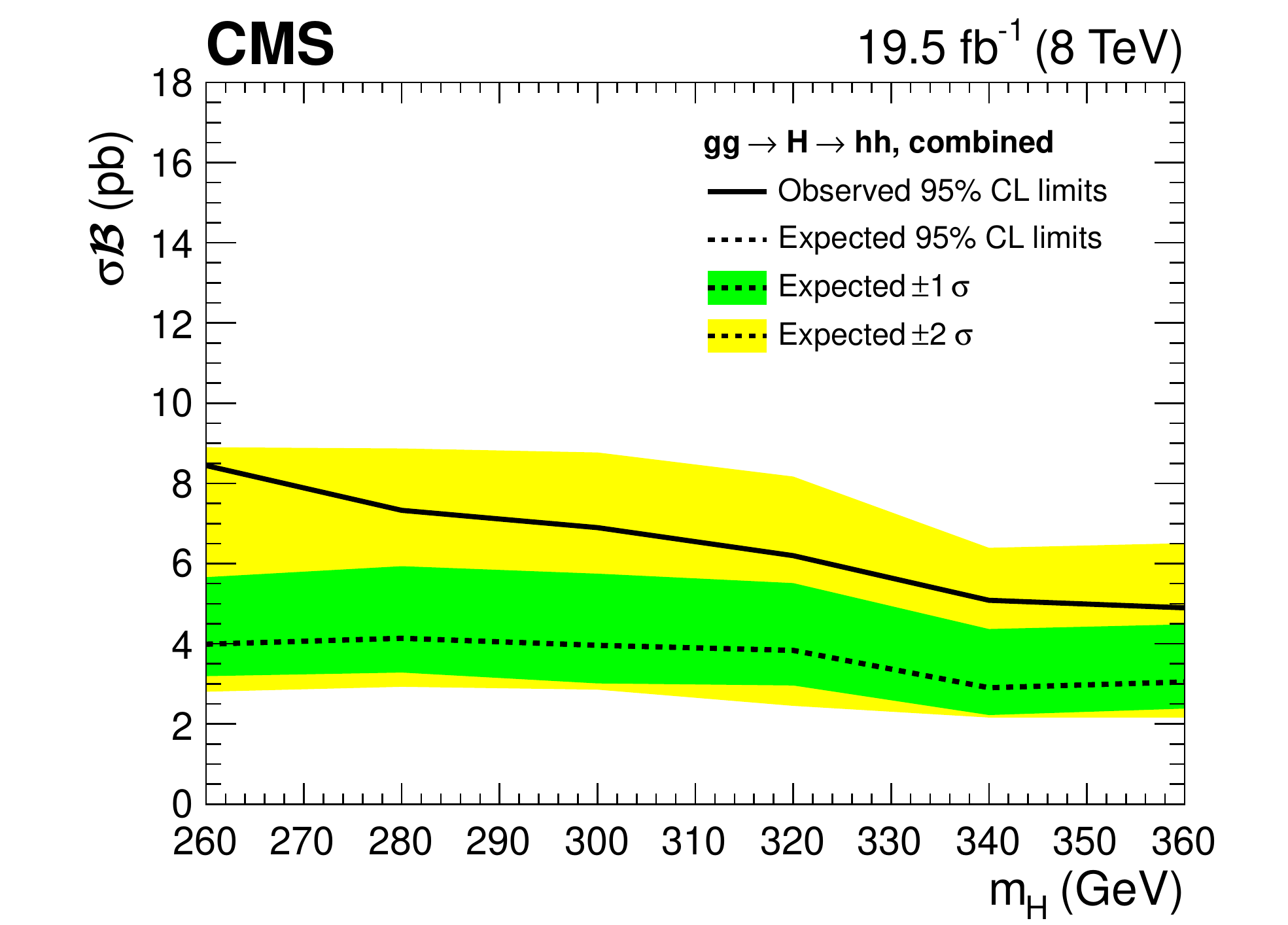}
\caption{\cmsLLeft: observed and expected 95\% \CL $\sigma \mathcal{B}$ limits for gluon fusion
production of $\PH$ and the decay $\PH \to \Ph \Ph$ with
one and two standard deviation bands shown. These limits are based only on multilepton channels.
The $\Ph$ branching fractions are assumed to have SM values.
\cmsRRight: the
same, but also including lepton+diphoton channels.}
\label{fig:HeavyHiggsLimitAll}
\end{figure}

Figure~\ref{fig:HeavyHiggsLimitAll}~(\cmsLeft) shows 95\% \CL observed and
expected $\sigma \mathcal{B}$ upper limit for the
gluon fusion production of heavy scalar $\PH$, with the decay
$\PH \to \Ph \Ph$
along with one and two standard deviation bands around the expected limits
using only the multilepton channels. Figure~\ref{fig:HeavyHiggsLimitAll}~(\cmsRight)
shows the same using both multilepton and diphoton channels. In placing these
model-independent limits, we explicitly assume that $\Ph$ is the recently discovered
SM-like Higgs boson~\cite{Aad:2012tfa,Chatrchyan:2012ufa,Chatrchyan:2013lba} particularly in regards to
the branching fraction of its various decay modes as predicted in the SM.

For low masses, there is an almost two standard deviation discrepancy between the
expected and observed 95\% \CL limits in Fig.~\ref{fig:HeavyHiggsLimitAll}. Its origin traces
back to three four-lepton channels listed in Table~\ref{tab:results4L}, which can also be located
in Fig.~\ref{fig:ResultsHhh}~(\cmsRight). They consist of events with a $\tauh$ and
three light leptons containing an off-$\cPZ$
OSSF dilepton pair, but not a $\cPqb$-tagged jet. The $\PH \to \Ph \Ph$ signal resides almost entirely
in the 0--100\GeV range in $\MET$ which is spanned by these three channels collectively.
The observed (expected) number of events is 11 (5.7 $\pm$ 1.7), 4 (2.4 $\pm$ 0.5)
and 5 (2.6 $\pm$ 0.6) for $\MET$ in ranges 0--30, 30--50, and 50--100\GeV,
respectively. Summing over the three channels, the observed count is 20 with an expectation of
10.7 $\pm$ 1.9, giving the probability of observing 20 events
over the 0--100\GeV $\MET$ range to be approximately 2.2\%.
Systematic uncertainties and their correlations are taken into
account when evaluating this probability.
The observed discrepancy in the limits shown in Fig.~\ref{fig:HeavyHiggsLimitAll} is thus consistent with a broad fluctuation
in the observed $\MET$ distribution.
Given the large number of channels
under scrutiny in this search, fluctuations at this level are to be expected.
No such deviation is observed in the $\MET$ distribution for other
search channels.

\begin{figure}[!ht]
\centering
\includegraphics[width=0.46 \textwidth]{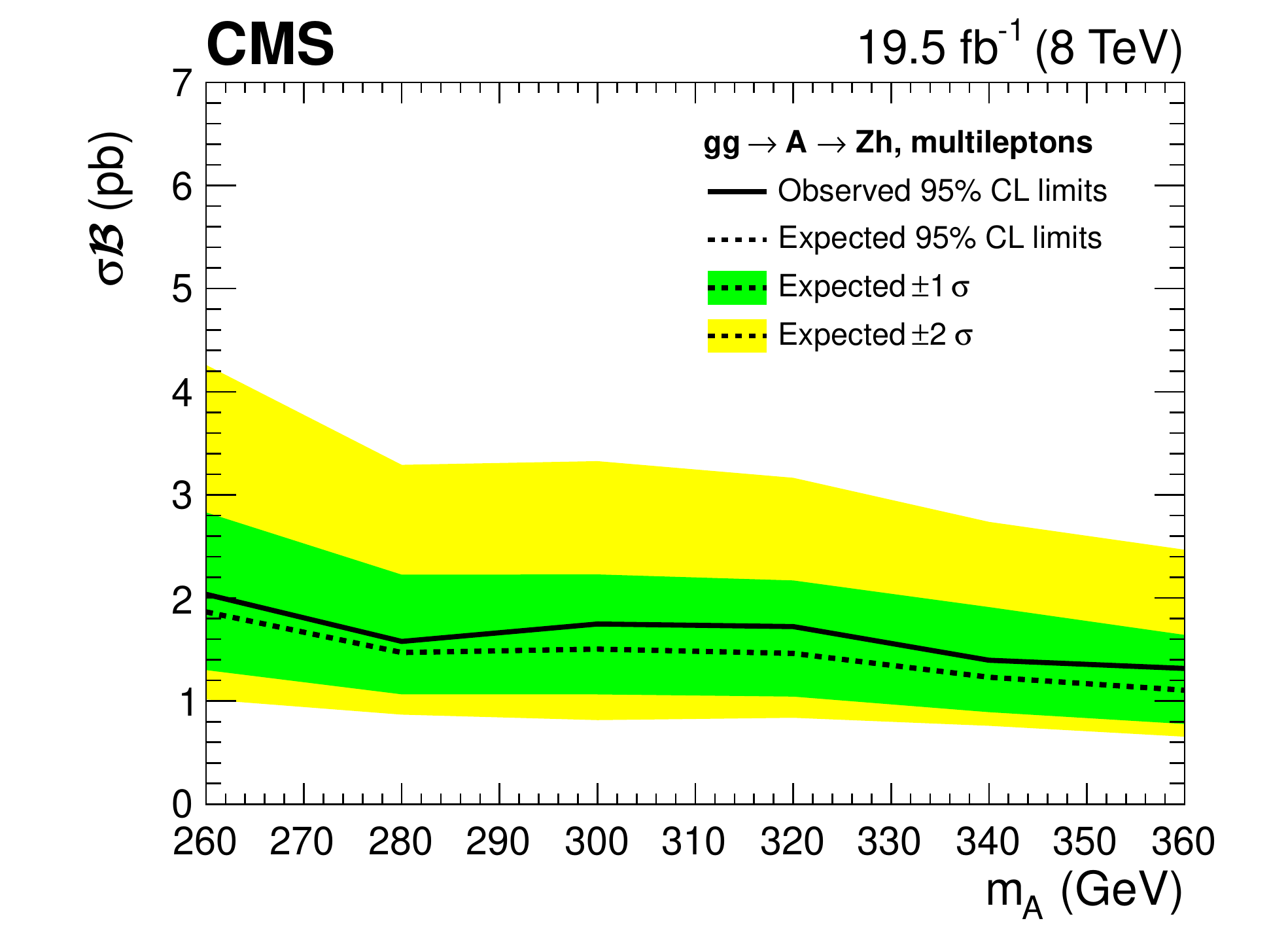}
\includegraphics[width=0.46 \textwidth]{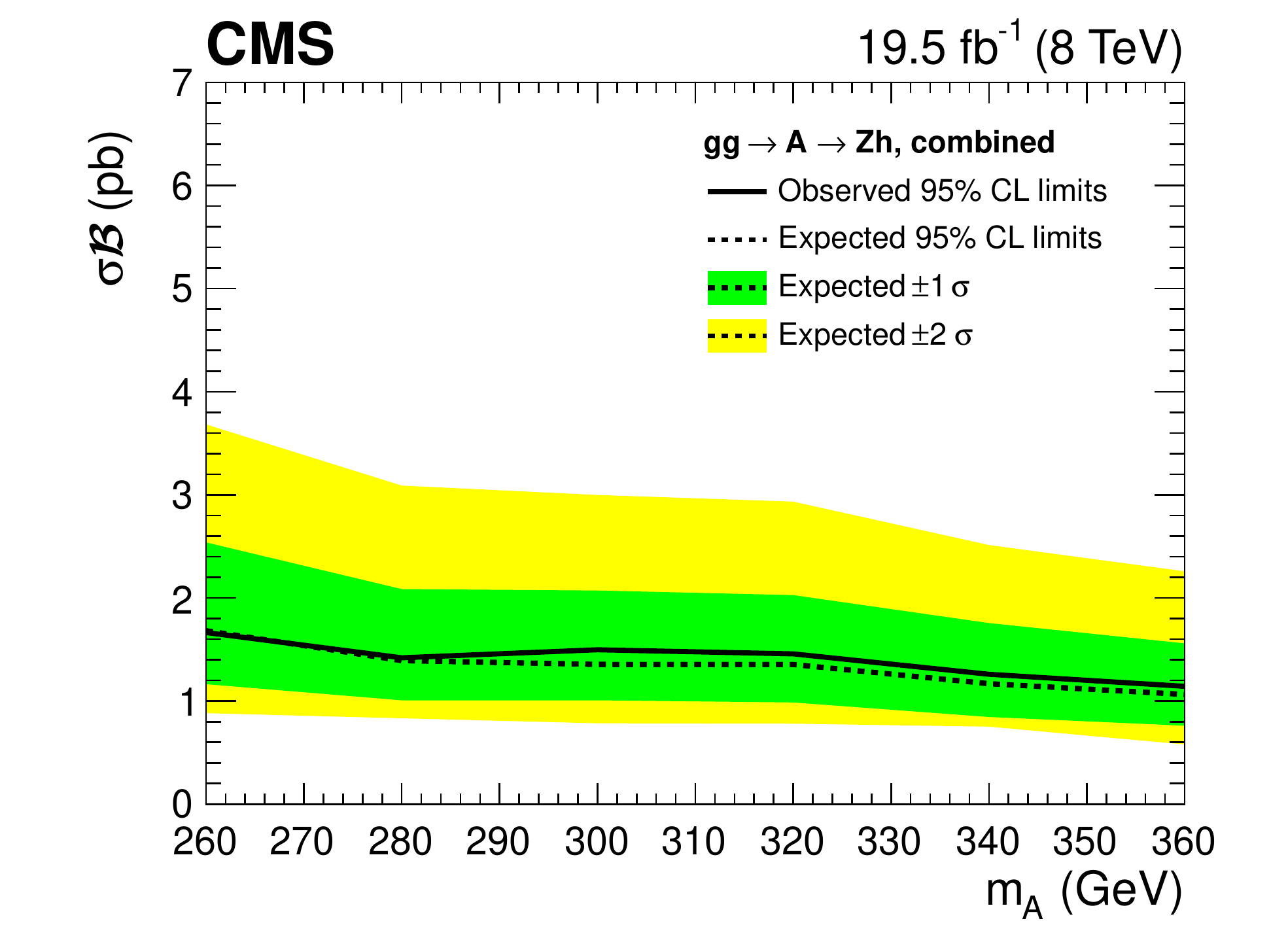}
\caption{\cmsLLeft: observed and expected 95\% \CL $\sigma \mathcal{B}$ limits for gluon fusion production of $\PA$ and the decay
$\PA \to \cPZ \Ph$
with one and two standard deviation bands shown. These limits are based only on multilepton channels.
The $\Ph$ branching fraction are assumed to have SM values.
\cmsRRight: the same, but also including lepton+diphoton channels.}
\label{fig:AtoZHiggsLimitAll}
\end{figure}

Next we probe the sensitivity to gluon fusion production of the heavy
pseudoscalar $\PA$  with the decay $\PA \to \cPZ \Ph$.
Figure~\ref{fig:AtoZHiggsLimitAll}~(left) shows 95\% \CL upper
limits on $\sigma \mathcal{B}$ for $\PA \to \cPZ \Ph$ search along with one
and two standard deviation bands around the expected contour using only the multilepton
channels. Figure~\ref{fig:AtoZHiggsLimitAll}~(right) shows the same signal
probed with both multilepton and diphoton channels.
The observed and expected exclusions are consistent.

\subsection{Interpretations in the context of two-Higgs-doublet models}

General models with two Higgs doublets may exhibit new tree-level contributions
to flavor-changing neutral currents that are strongly constrained by low-energy experiments.
Prohibitive flavor violation is avoided in a 2HDM if all fermions of a given representation receive
their masses through renormalizable Yukawa couplings to a single Higgs doublet, as in the
case of supersymmetry. There are four such possible distinct assignments of fermion couplings
in models with two Higgs doublets, the most commonly considered of which are called Type I and Type II models.
In Type I models all fermions receive their masses through Yukawa couplings to a single Higgs doublet,
while in Type II models the up-type quarks receive their masses through couplings to one doublet
and down-type quarks and leptons couple to the second doublet. In either type, after electroweak
symmetry breaking the physical Higgs scalars are linear combinations of these two electroweak
Higgs doublets, so that fermion couplings to the physical states depend on the type of 2HDM, the
mixing angle $\alpha$, and the ratio of vacuum expectation values $\tan \beta$.
We next present search interpretations in the context of Type I and
Type II 2HDMs~\cite{Craig:2012vn}. In these models,
the production cross sections for $\PH$ and $\PA$ as well as the branching
fractions for them to decay to two SM-like Higgs bosons depend on parameters
$\alpha$ and $\tan \beta$. The mixing angle between $\PH$ and $\Ph$ is
given by $\alpha$, whereas $\tan \beta$ gives the relative contribution
of each Higgs doublet to electroweak symmetry breaking. In obtaining
these model-dependent limits, the daughter $\Ph$ is assumed to
be the recently discovered SM-like Higgs boson, but the
branching fractions to its various decay modes are assumed to
be dictated by the parameters $\alpha$ and $\tan \beta$ of the 2HDM,
as described below.

\begin{figure*}[!ht]
\centering
\includegraphics[width=0.46 \textwidth]{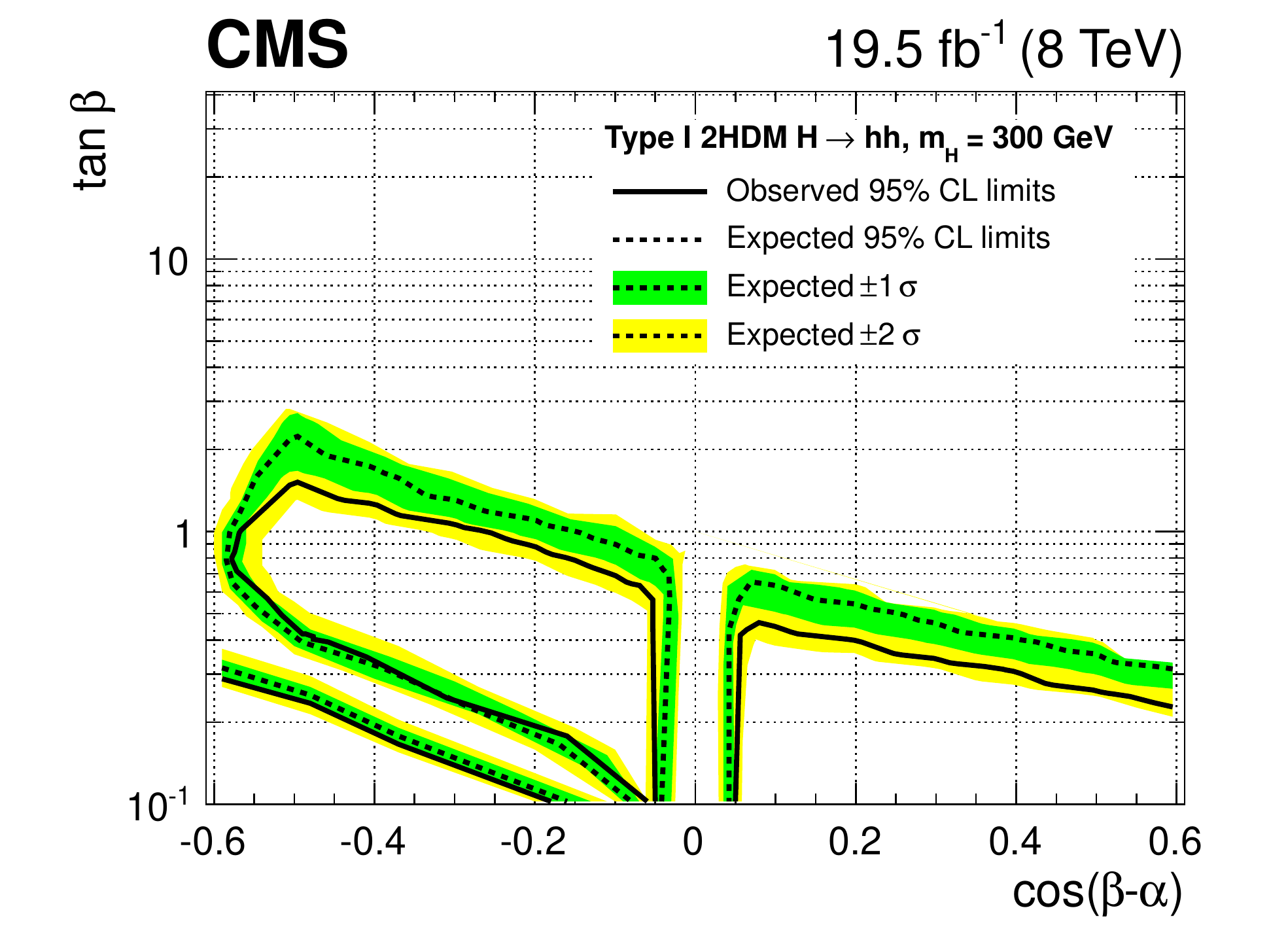}
\includegraphics[width=0.415 \textwidth]{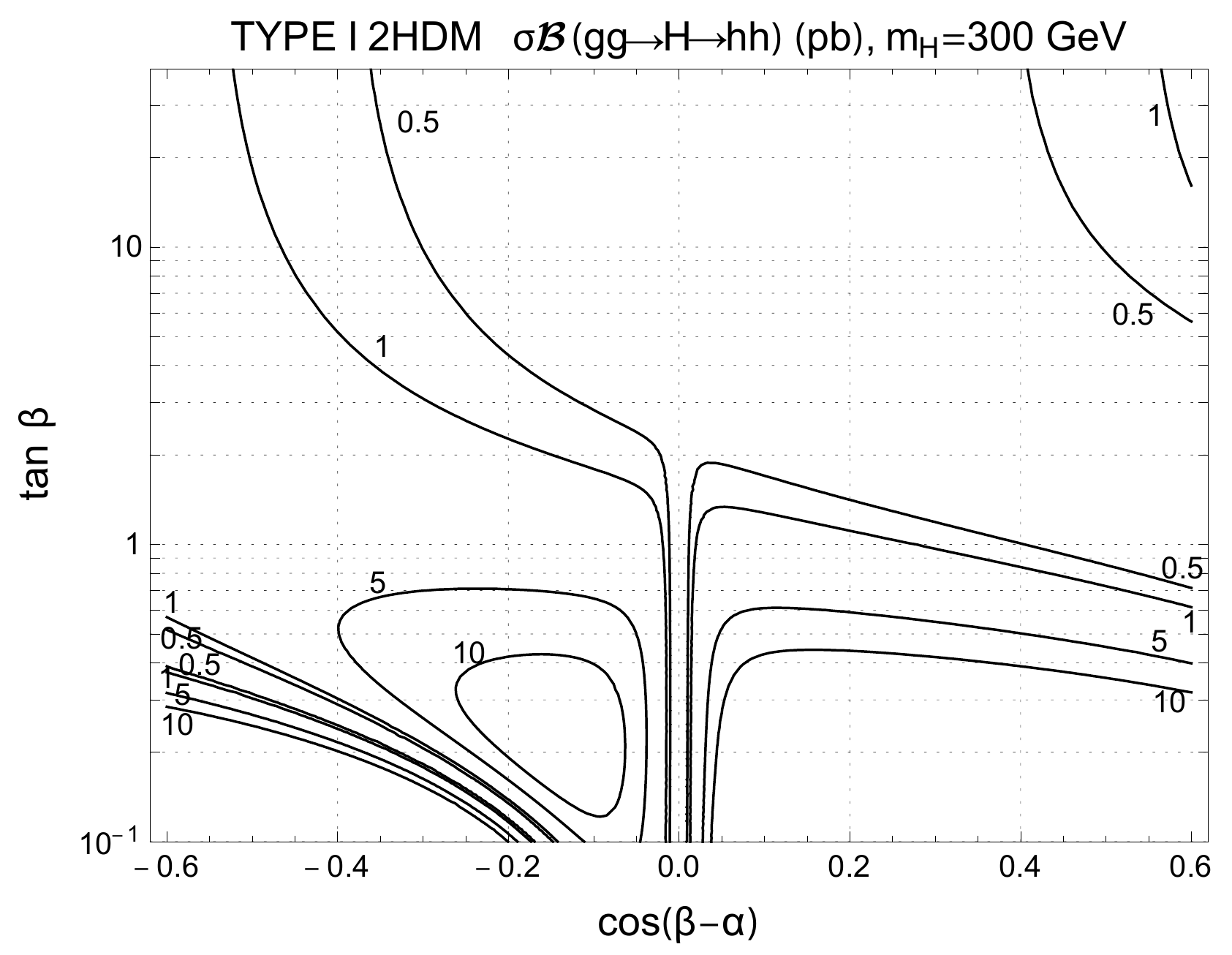}
\includegraphics[width=0.46 \textwidth]{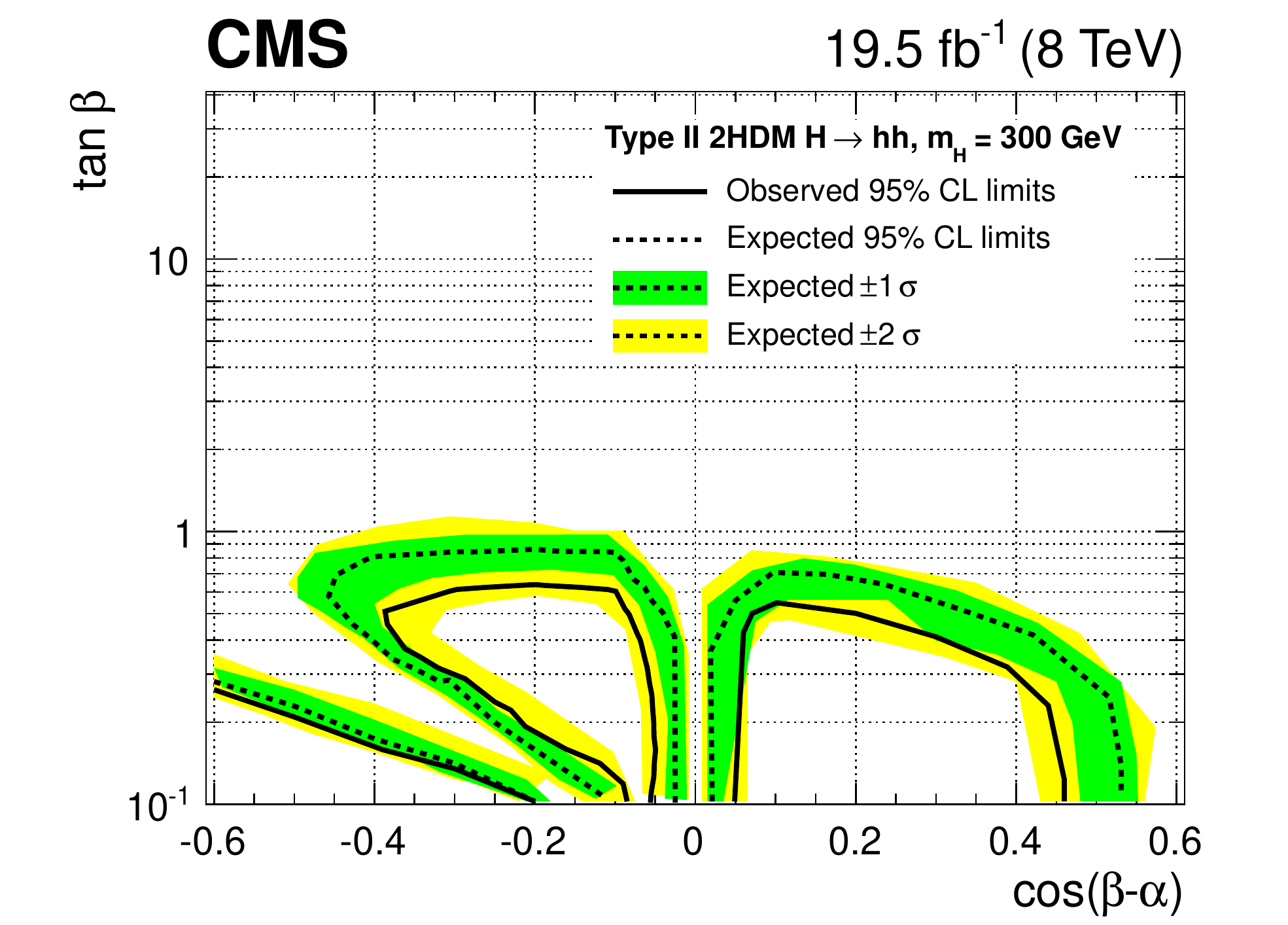}
\includegraphics[width=0.415 \textwidth]{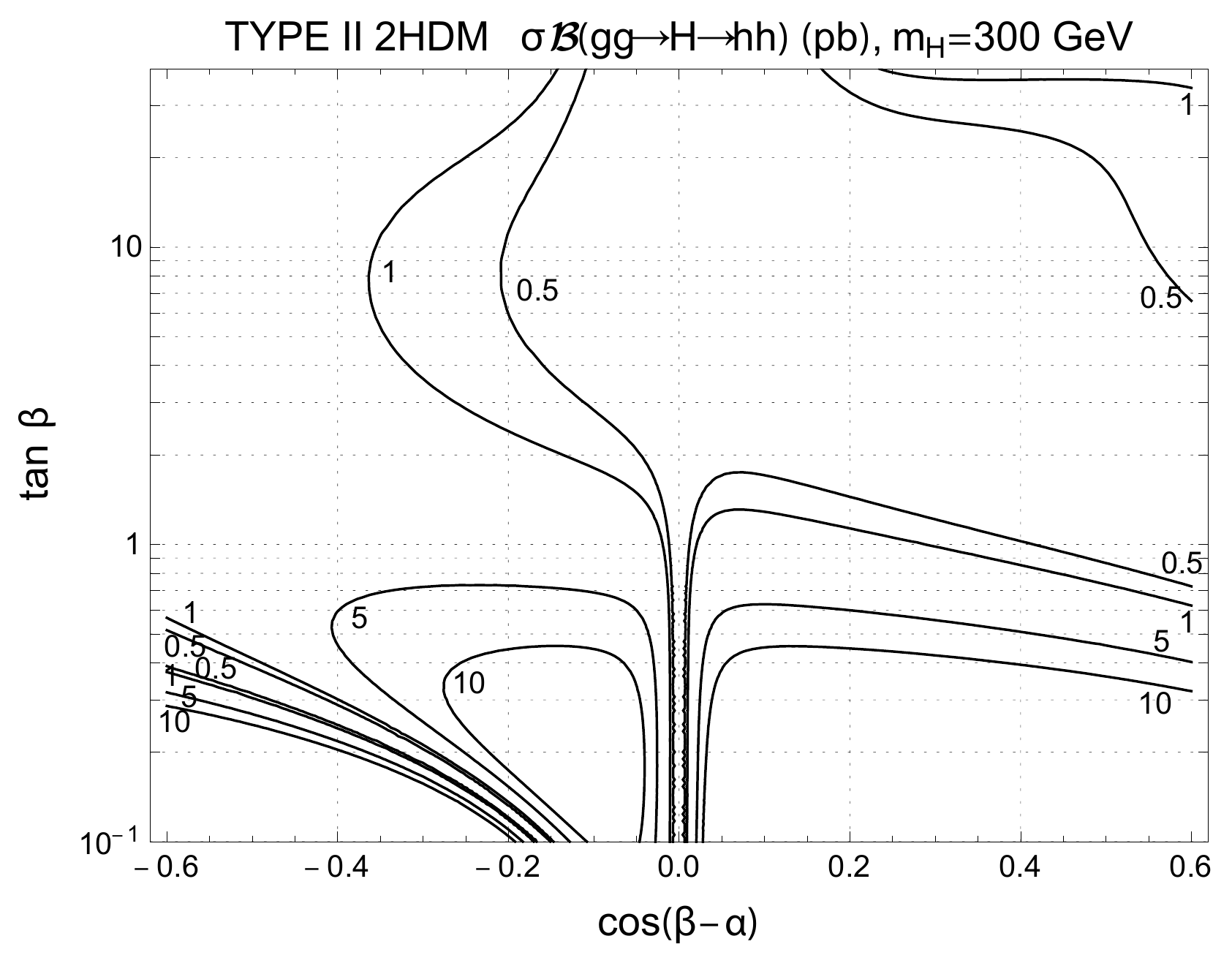}
\caption{Left: observed and expected 95\% \CL upper limits for gluon
fusion production of a heavy Higgs boson
$\PH$ of mass 300\GeV as a function of parameters $\tan \beta$ and $\cos(\beta - \alpha)$ of the Type I
(upper) and II (lower) 2HDM. The parameters determine the $\PH$
production cross section as well as the branching fractions
$\mathcal{B}(\PH \to \Ph \Ph)$ and
$\mathcal{B}(\Ph \to \PW \PW^*, \cPZ \cPZ^*, \Pgt \Pgt, \gamma \gamma)$,
which are relevant to this search. Right: the $\sigma \mathcal{B}(\PH \to \Ph \Ph)$
contours for Type I (upper) and II (lower) 2HDM adopted from Ref.~\cite{Craig:2013hca}.
The excluded regions are either below the open limit contours or within the closed ones.}
\label{fig:T1Hhh0new2hdmAndT2Hhh0new2hdm}
\end{figure*}

\begin{figure*}[!ht]
\centering
\includegraphics[width=0.46 \textwidth]{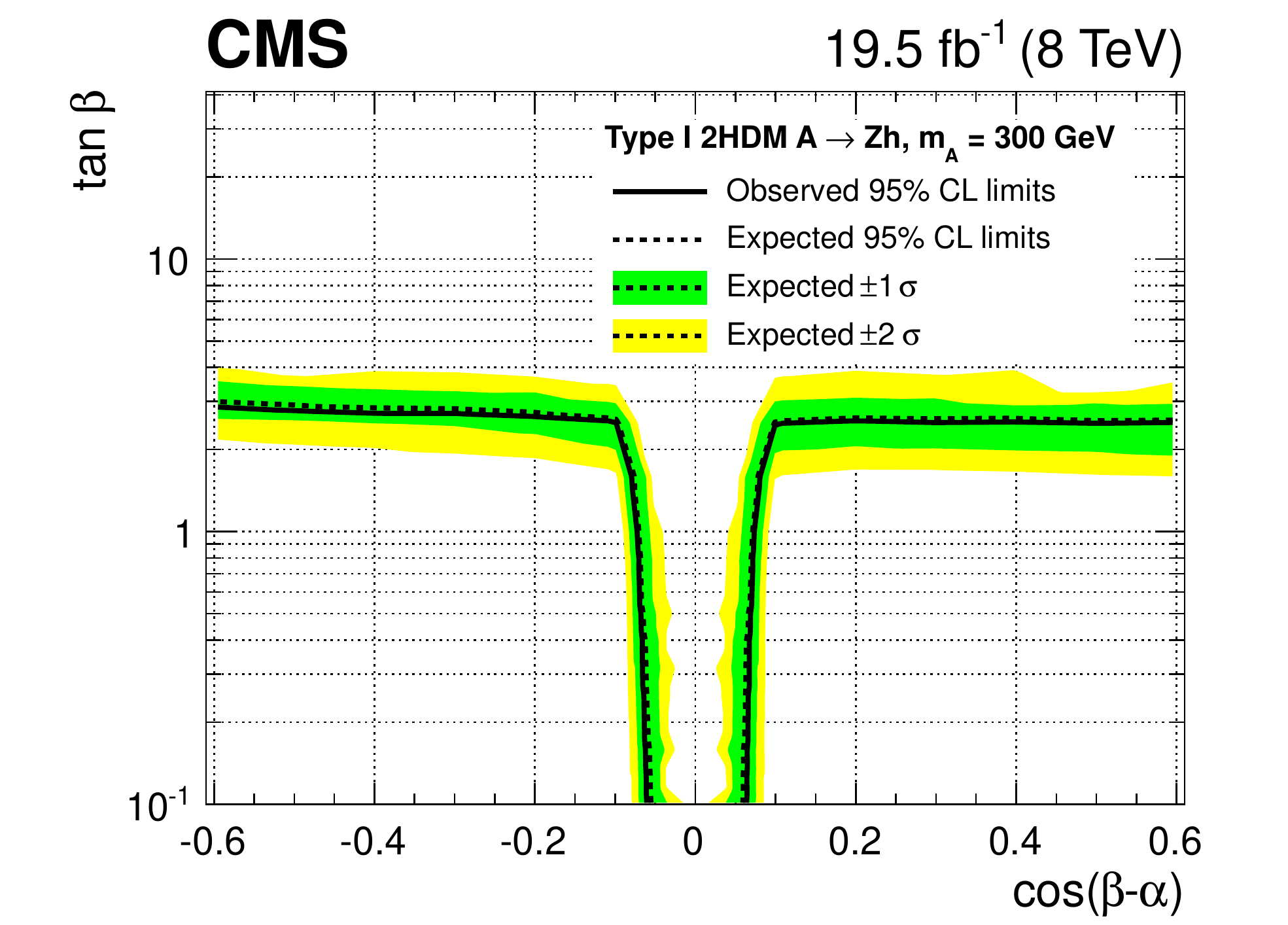}
\includegraphics[width=0.415 \textwidth]{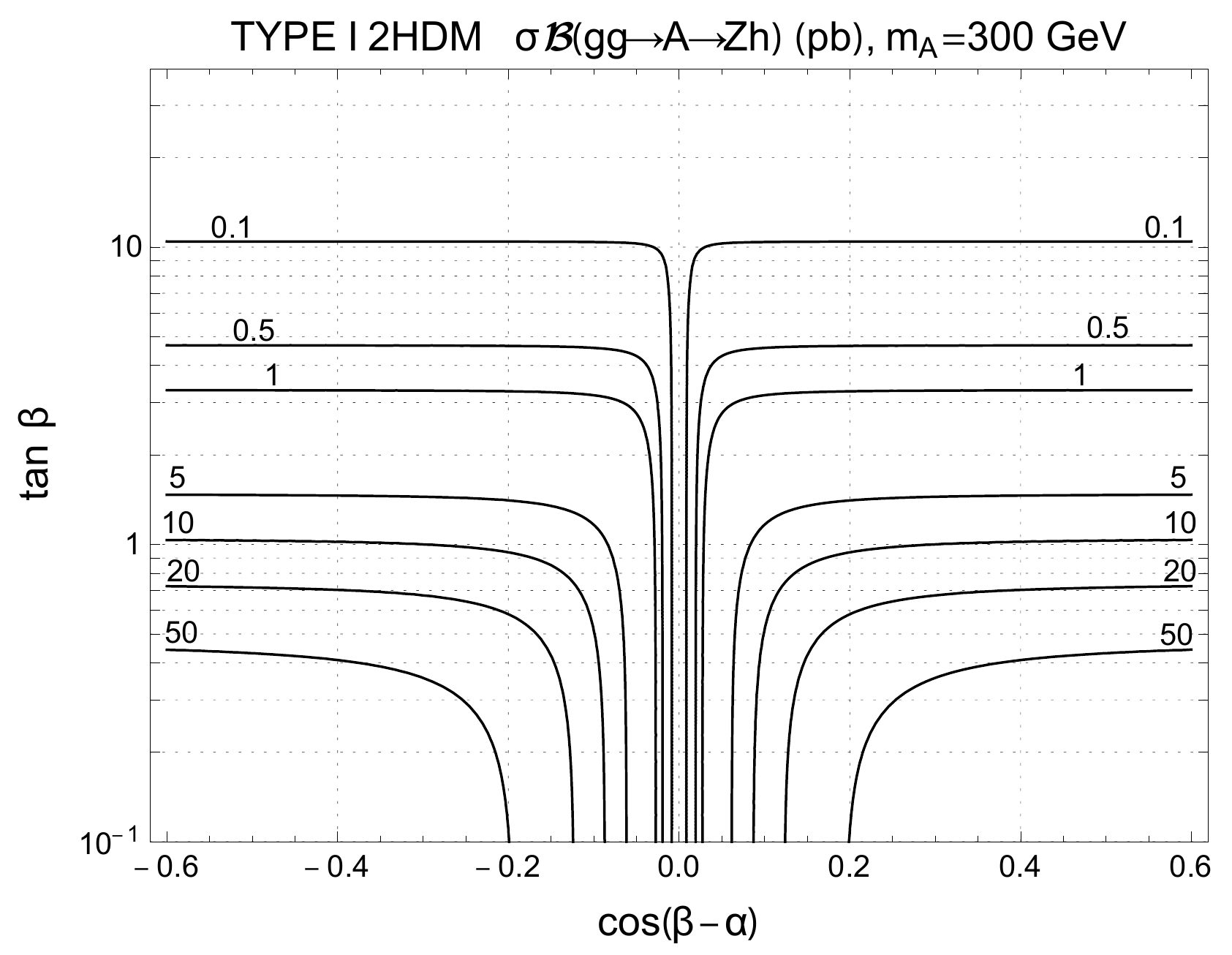}
\includegraphics[width=0.46 \textwidth]{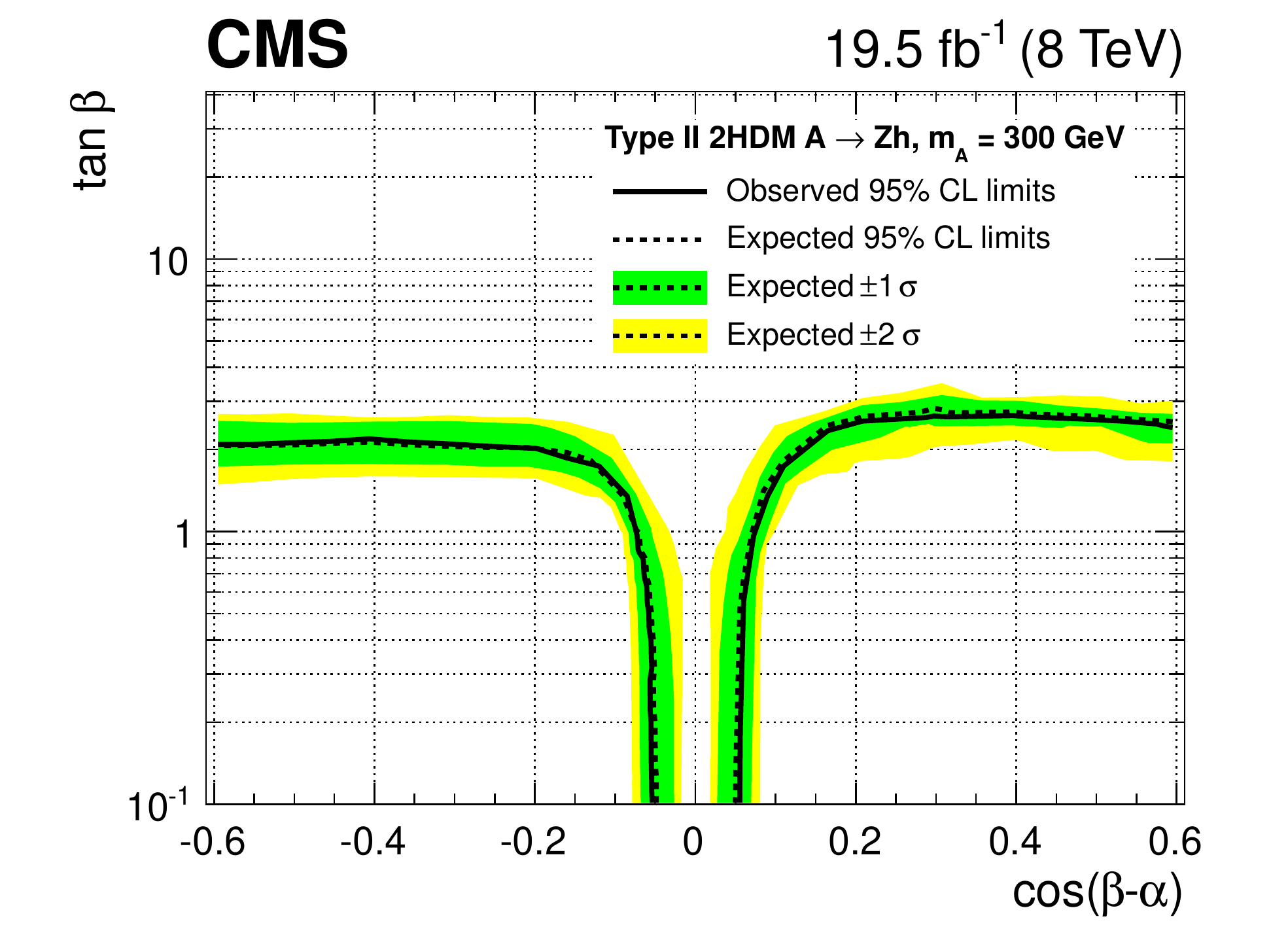}
\includegraphics[width=0.415 \textwidth]{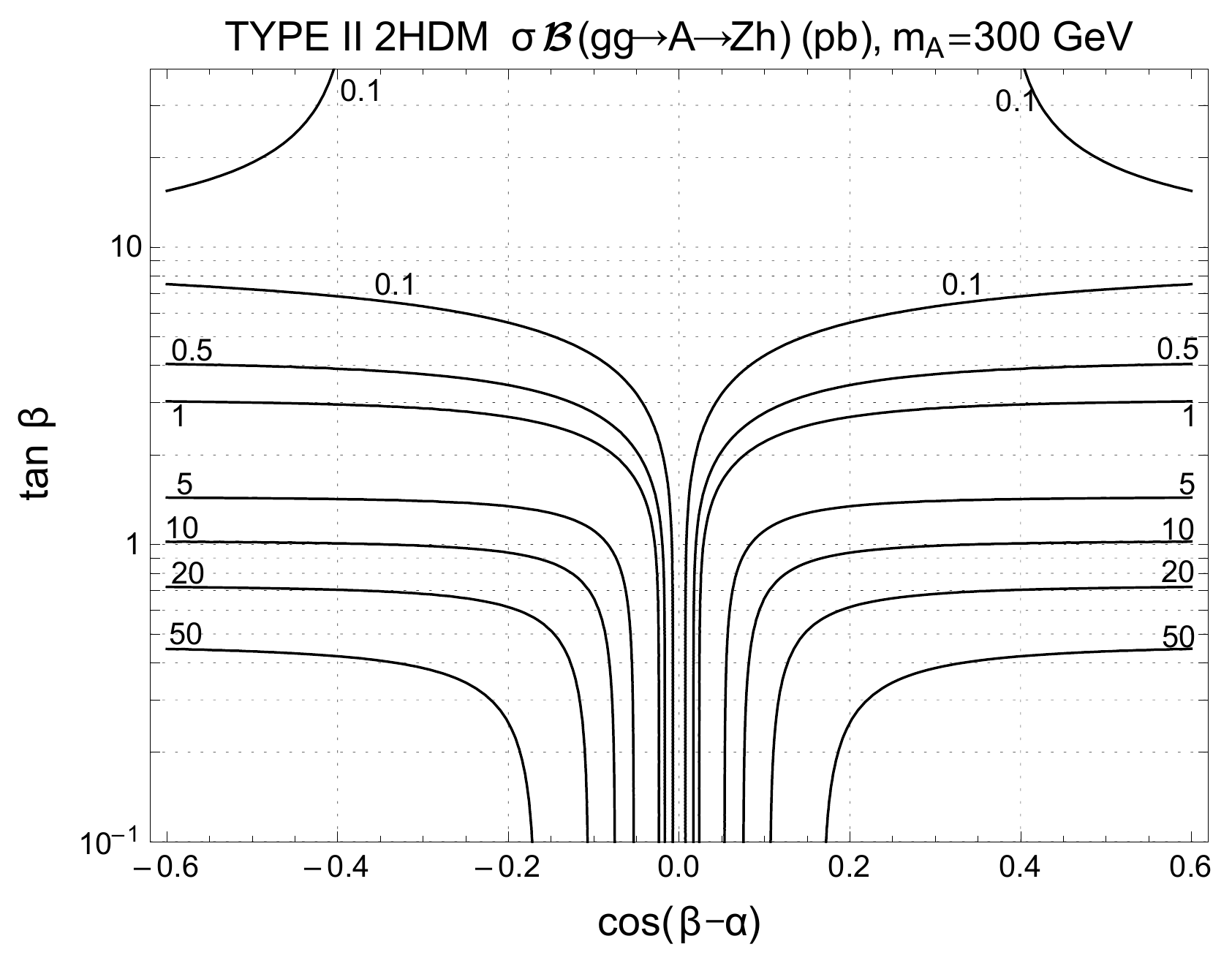}
\caption{Left: observed and expected 95\% \CL upper limits for
gluon fusion production of an $\PA$ boson
of mass 300\GeV as a function of parameters $\tan \beta$ and $\cos(\beta - \alpha)$ of the
Type I (upper) and II (lower) 2HDM. The parameters determine the $\PA$ production
cross section as well as the branching fractions $\mathcal{B}(\PA \to \cPZ \Ph)$ and
$\mathcal{B}(\Ph \to \PW \PW^*, \cPZ \cPZ^*, \Pgt \Pgt, \gamma \gamma)$ which
are relevant to this search. Right: the $\sigma \mathcal{B}(\PA \to \cPZ \Ph)$
contours for Type I (upper) and II (lower) 2HDM adopted from Ref.~\cite{Craig:2013hca}.
The excluded regions are below the open limit contours.}
\label{fig:T1AZhnew2hdmAndT2AZhnew2hdm}
\end{figure*}

We use the \textsc{SusHi}~\cite{Harlander:2012pb} program to
obtain the 2HDM cross sections. The branching fraction for SM-like Higgs boson
are calculated using the \textsc{2hdmc}~\cite{Eriksson:2009ws} program.
The \textsc{2hdmc} results are consistent with those provided by the LHC Higgs
cross section working group~\cite{Dittmaier:2011ti}.  A detailed list of couplings
of $\PH$ and $\PA$ to SM fermions and massive gauge bosons in Type I and II 2HDMs
has been tabulated in Ref.~\cite{Craig:2012pu}.
Figure~\ref{fig:T1Hhh0new2hdmAndT2Hhh0new2hdm}~(top left and bottom left) shows observed
and expected 95\% \CL upper limits
for gluon fusion production of
a heavy Higgs boson $\PH$ of mass 300\GeV for Type I and Type II 2HDMs,
respectively, along with the $\sigma \mathcal{B}$ theoretical predictions~(right)
adopted from Ref.~\cite{Craig:2013hca} for the two models.
Figure~\ref{fig:T1AZhnew2hdmAndT2AZhnew2hdm}~(top left and bottom left)
shows similar results for the pseudoscalar $\PA$ Higgs boson of mass
300\GeV.
The branching fractions for
the SM-like Higgs boson daughters of the $\PH$ and $\PA$ vary across
the $\tan \beta$ versus $\cos(\beta-\alpha)$ plane and are incorporated in the upper
limit calculations.

\begin{figure}[!ht]
\centering
\includegraphics[width=0.46 \textwidth]{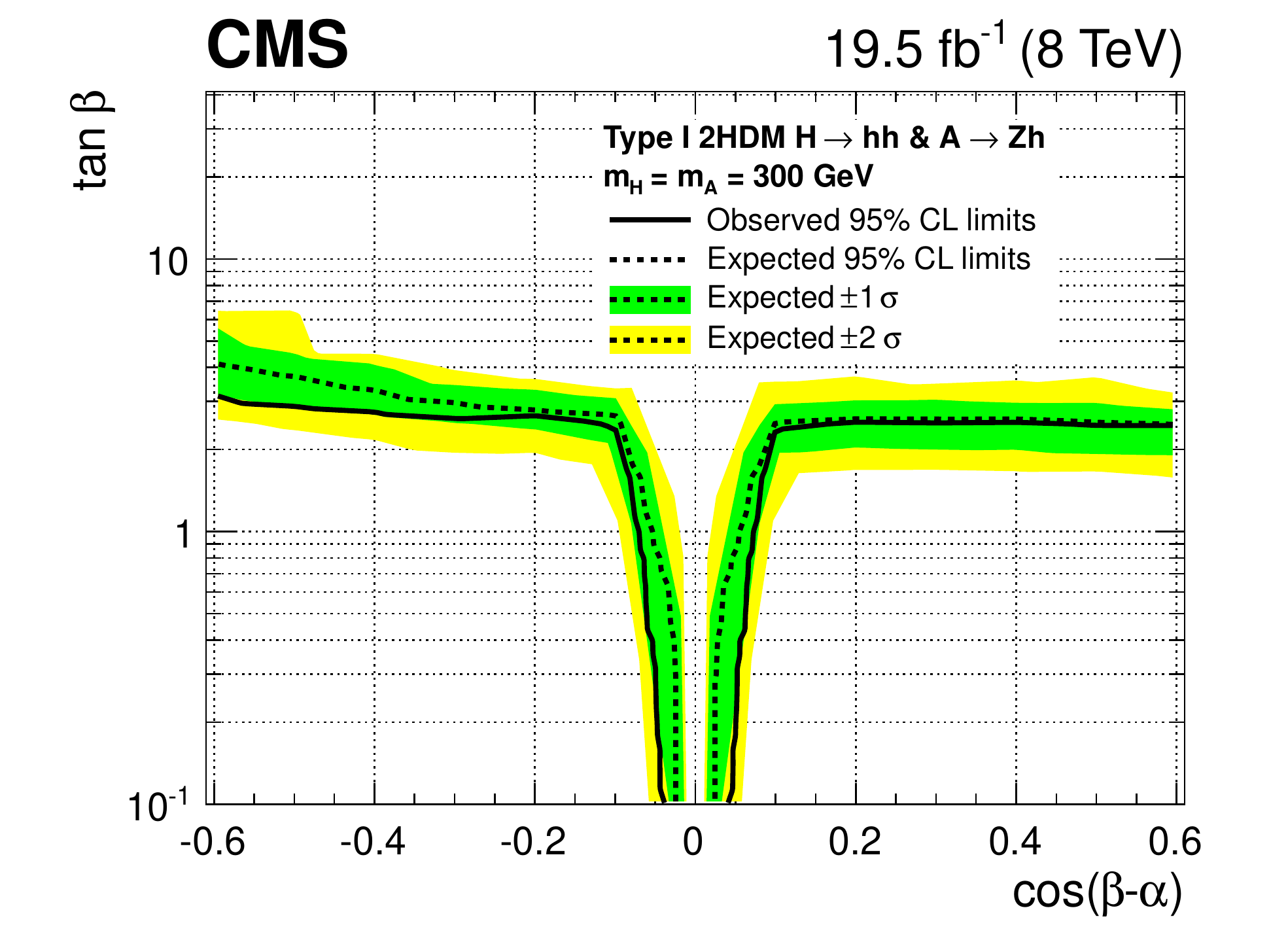}
\includegraphics[width=0.46 \textwidth]{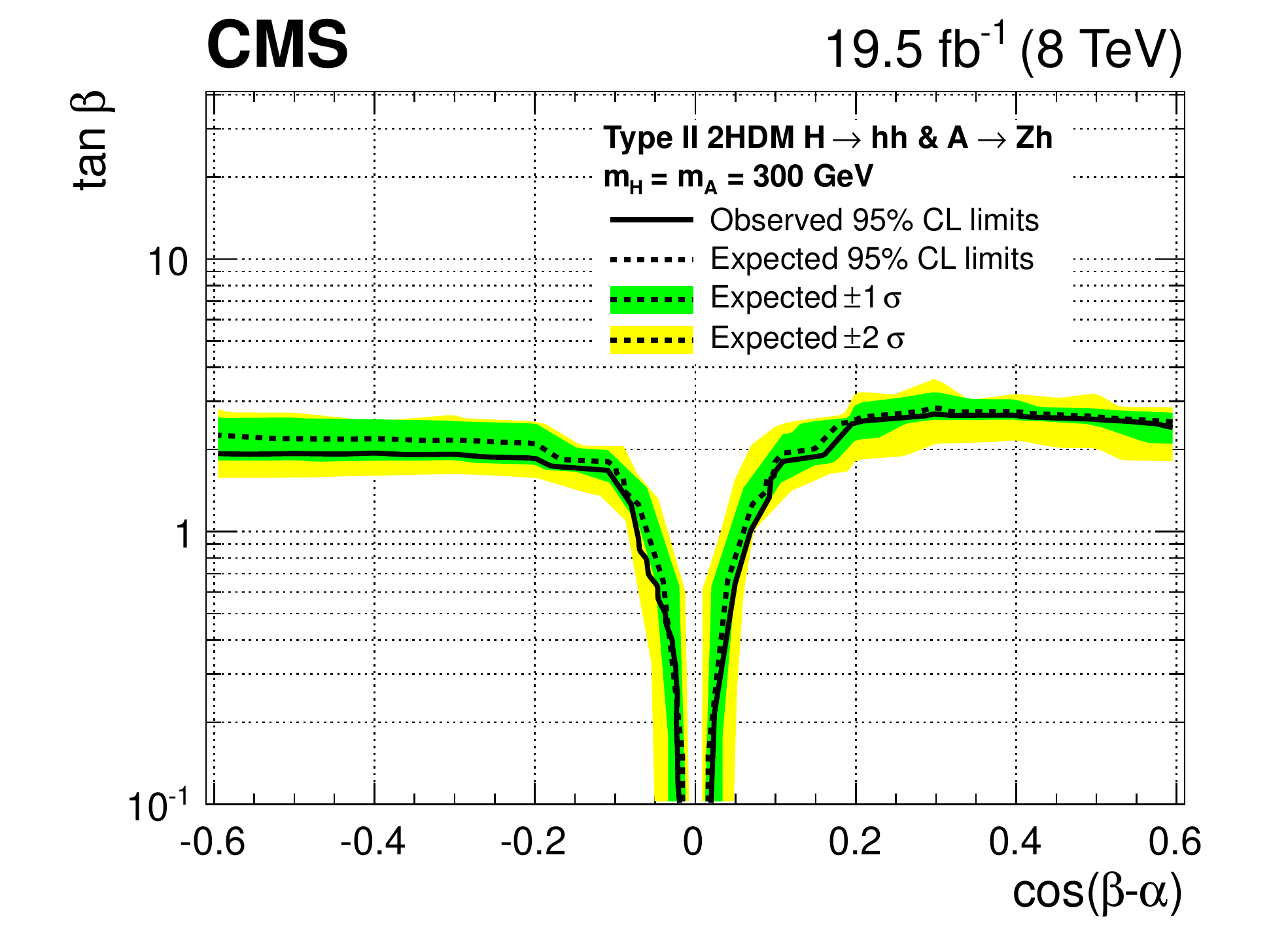}			
\caption{Combined observed and expected 95\% upper limits for gluon fusion
production of a heavy Higgs
boson $\PH$ and $\PA$ of mass 300\GeV for Type I (\cmsLeft) and Type II~(\cmsRight) 2HDM
as a function of parameters $\tan \beta$ and $\cos(\beta - \alpha)$. The parameters determine
the $\PH$ and $\PA$ production cross sections as well as the branching fractions
$\mathcal{B}(\PH \to \Ph \Ph)$,
$\mathcal{B}(\PA \to \cPZ \Ph)$,
and
$\mathcal{B}(\Ph \to \PW \PW^*, \cPZ \cPZ^*, \Pgt \Pgt, \gamma \gamma)$,
which are relevant to this search.}
\label{fig:T1HandAcombined}
\end{figure}

Finally, we further improve constraints on the 2HDM parameters using the simultaneous
null findings for the $\PH$ and $\PA$. Figure~\ref{fig:T1HandAcombined} shows exclusion in
$\tan \beta$ versus $\cos(\beta - \alpha)$ plane for the combined gluon
fusion signal for Type I (\cmsLeft)
and Type II~(\cmsRight) 2HDMs, assuming $\PH$ and $\PA$ to be mass degenerate
with a mass of 300\GeV. Once again,
the branching fractions of the SM-like $\Ph$ daughters are allowed to vary across the plane.

\subsection{\texorpdfstring{$\cPqt \to \cPqc \Ph$}{t -> c h} search results}

The $\cPqt \to \cPqc \Ph$ signal predominantly populates lepton+diphoton channels
with a $\cPqb$-tag and $\ell \ell \ell$ (no $\tauh$) multilepton channels that lack an OSSF
dilepton pair or have an off-$\cPZ$ OSSF pair. Beyond the fact that the presence of
charm quark increases the likelihood of an event being classified as being $\cPqb$-tagged,
no special effort is made to identify the charm quark present in the signal.
The observations and SM expectations
for the ten most sensitive channels are listed in Table~\ref{tab:tchChannels} along
with the signal yield for a nominal value of $\mathcal{B}(\cPqt \to \cPqc \Ph) = 1\%$.
No significant excess is observed.

\begin{table*}[!htbp]
\centering
\topcaption{The ten most sensitive search channels for $\cPqt \to \cPqc \Ph$, along
with the number of observed~(Obs.), expected SM background~(Exp.), and expected
signal~(Sig.) events (assuming $\mathcal{B}(\cPqt \to \cPqc \Ph) = 1\%$). The three-lepton
channels are taken from Ref.~\cite{Chatrchyan:2014aea}, have $\HT < 200\GeV$
and do not contain a $\tauh$. The stated uncertainties contain
both systematic and statistical components.}
\label{tab:tchChannels}
\begin{scotch}{c|c|c|c|c|c}
Channel & $\MET$ (\GeVns{}) & ${N}_{\cPqb}$ & Obs. & Exp. & Sig. \\
\hline
\multirow{6}{*}{$\gamma \gamma \ell$}
& (50, 100) & ${\ge}1$ & 1 & 2.3 $\pm$ 1.2 & 2.88 $\pm$ 0.39 \\
& (30, 50) & ${\ge}1$ & 2 & 1.1 $\pm$ 0.6 & 2.16 $\pm$ 0.30 \\
& (0, 30) & ${\ge}1$ & 2 & 2.1 $\pm$ 1.1 & 1.76 $\pm$ 0.24 \\
& (50, 100) & 0 & 7 & 9.5 $\pm$ 4.4 & 2.22 $\pm$ 0.31 \\
& (100, $\infty$) & ${\ge}1$ & 0 & 0.5 $\pm$ 0.4 & 0.92 $\pm$ 0.14 \\
& (100, $\infty$) & 0 & 1 & 2.2 $\pm$ 1.0 & 0.94 $\pm$ 0.17 \\
\hline
\multirow{2}{*}{$\genfrac{}{}{0pt}{0}{\ell \ell \ell}{\text{(OSSF1, below-$\cPZ$)}}$} & (50, 100) & ${\ge}1$ & 48 & 48 $\pm$ 23 & 9.5 $\pm$ 2.3 \\
& (0, 50) & ${\ge}1$ & 34 & 42 $\pm$ 11 & 5.9 $\pm$ 1.2 \\
\hline
\multirow{2}{*}{$\genfrac{}{}{0pt}{0}{\ell \ell \ell}{\text{(OSSF0)}}$} & (50, 100) & ${\ge}1$ & 29 & 26 $\pm$ 13 & 5.9 $\pm$ 1.3 \\
& (0, 50) & ${\ge}1$ & 29 & 23 $\pm$ 10 & 4.3 $\pm$ 1.1 \\
\end{scotch}
\end{table*}

The statistical procedure yields an observed limit of $\mathcal{B}(\cPqt \to \cPqc \Ph) = 0.56\%$
and an expected limit of $\mathcal{B}(\cPqt \to \cPqc \Ph) = (0.65^{+0.29}_{-0.19})\%$ from SM $\ttbar$ production followed by either $\cPqt \to \cPqc \Ph$
or its charge-conjugate decay.
The $\cPqt \to \cPqc \Ph$ branching fraction is related to the left- and right-handed top
flavor changing Yukawa couplings $\lambda_{\cPqt \cPqc}^{\Ph}$ and $\lambda_{\cPqc \cPqt}^{\Ph}$, respectively,
by $\mathcal{B}(\cPqt \to \cPqc \Ph) \simeq 0.29 \left(\abs{\lambda_{\cPqt \cPqc}^{\Ph}}^2 + \abs{\lambda_{\cPqc \cPqt}^{\Ph}}^2 \right)$
~\cite{Craig:2012vj}, so that the observed limit corresponds to a limit on the
couplings of $\sqrt{\abs{\lambda_{\cPqt \cPqc}^{\Ph}}^2 + \abs{\lambda_{\cPqc \cPqt}^{\Ph}}^2 } < 0.14$.

To facilitate interpretations in broader contexts~\cite{Chen:2013qta},
we also provide limits on $\mathcal{B}(\cPqt \to \cPqc \Ph)$ from
individual Higgs boson decay modes. For this purpose, we assume the SM
branching fraction for the
Higgs boson decay mode~\cite{Chatrchyan:2013mxa} under consideration,
and ignore other decay modes. Table~\ref{tab:tch} shows the results,
illustrating the analysis sensitivity for the $\cPqt \to \cPqc \Ph$
decay in each of the Higgs boson decay modes.

\begin{table*}[h!tb]
\centering
\topcaption{Comparison of the observed and expected 95\% \CL limits
on $\mathcal{B}(\cPqt \to \cPqc \Ph)$ from individual Higgs boson decay modes along with
the 68\% \CL uncertainty ranges.}
\label{tab:tch}
\begin{scotch}{ll|ccc}
\multicolumn{2}{c|}{Higgs boson decay mode} & \multicolumn{3}{c}{Upper limits on $\mathcal{B}(\cPqt \to \cPqc \Ph)$} \\
& & Obs. & Exp. & 68\% \CL range \\
\hline
$\mathcal{B}(\Ph \to \PW \PW^*)$ & ${=}~23.1\%$  & 1.58\% & 1.57\% & (1.02--2.22)\% \\
$\mathcal{B}(\Ph \to \Pgt \Pgt)$ & ${=}~6.15\%$& 7.01\% & 4.99\% & (3.53--7.74)\% \\
$\mathcal{B}(\Ph \to \cPZ \cPZ^*)$ & ${=}~2.89\%$ & 5.31\% & 4.11\% & (2.85--6.45)\% \\
\hline
\multicolumn{2}{c|}{Combined multileptons ($\PW \PW^*$, $\Pgt\Pgt$, $\cPZ \cPZ^*$)} & 1.28\% & 1.17\% & (0.85--1.73)\% \\
\hline
$\mathcal{B}(\Ph \to \gamma \gamma)$ & ${=}~0.23\%$ & 0.69\% & 0.81\% & (0.60--1.17)\% \\
\hline
\multicolumn{2}{c|}{Combined multileptons + diphotons} & 0.56\% & 0.65\% & (0.46--0.94)\% \\
\end{scotch}
\end{table*}

\section{Summary}
\label{concl}

We have performed a search for the $\PH \to \Ph \Ph$ and $\PA \to \cPZ \Ph$ decays
of heavy scalar ($\PH$) and pseudoscalar ($\PA$) Higgs bosons,
respectively, which occur in the extended Higgs sector described by the 2HDM.
This is the first search for these decays carried out at the LHC.
We used multilepton and diphoton final states from a dataset corresponding to an
integrated luminosity of $\procLumi$ of data recorded in 2012 from $\Pp \Pp$ collisions
at a center-of-mass energy of 8\TeV. We find no significant deviation from
the SM expectations and place 95\% \CL cross section upper limits of
approximately 7\unit{pb} on $\sigma \mathcal{B}$ for $\PH \to \Ph \Ph$ and 2\unit{pb} for $\PA \to \cPZ \Ph$.
We further interpret these limits in the context of Type I and Type II 2HDMs, presenting
exclusion contours in the $\tan \beta$ versus $\cos(\beta - \alpha$) plane.

Using diphoton and multilepton search channels that are sensitive to
the decay $\cPqt \to \cPqc \Ph$, we place an upper limit of 0.56\%
on $\mathcal{B}(\cPqt \to \cPqc \Ph)$, where the expected limit is
0.65\%. This is a significant improvement over the earlier limit of
1.3\% from the multilepton search alone~\cite{Chatrchyan:2014aea}.

\begin{acknowledgments}
\hyphenation{Bundes-ministerium Forschungs-gemeinschaft Forschungs-zentren} We congratulate our colleagues in the CERN accelerator departments for the excellent performance of the LHC and thank the technical and administrative staffs at CERN and at other CMS institutes for their contributions to the success of the CMS effort. In addition, we gratefully acknowledge the computing centers and personnel of the Worldwide LHC Computing Grid for delivering so effectively the computing infrastructure essential to our analyses. Finally, we acknowledge the enduring support for the construction and operation of the LHC and the CMS detector provided by the following funding agencies: the Austrian Federal Ministry of Science, Research and Economy and the Austrian Science Fund; the Belgian Fonds de la Recherche Scientifique, and Fonds voor Wetenschappelijk Onderzoek; the Brazilian Funding Agencies (CNPq, CAPES, FAPERJ, and FAPESP); the Bulgarian Ministry of Education and Science; CERN; the Chinese Academy of Sciences, Ministry of Science and Technology, and National Natural Science Foundation of China; the Colombian Funding Agency (COLCIENCIAS); the Croatian Ministry of Science, Education and Sport, and the Croatian Science Foundation; the Research Promotion Foundation, Cyprus; the Ministry of Education and Research, Estonian Research Council via IUT23-4 and IUT23-6 and European Regional Development Fund, Estonia; the Academy of Finland, Finnish Ministry of Education and Culture, and Helsinki Institute of Physics; the Institut National de Physique Nucl\'eaire et de Physique des Particules~/~CNRS, and Commissariat \`a l'\'Energie Atomique et aux \'Energies Alternatives~/~CEA, France; the Bundesministerium f\"ur Bildung und Forschung, Deutsche Forschungsgemeinschaft, and Helmholtz-Gemeinschaft Deutscher Forschungszentren, Germany; the General Secretariat for Research and Technology, Greece; the National Scientific Research Foundation, and National Innovation Office, Hungary; the Department of Atomic Energy and the Department of Science and Technology, India; the Institute for Studies in Theoretical Physics and Mathematics, Iran; the Science Foundation, Ireland; the Istituto Nazionale di Fisica Nucleare, Italy; the Korean Ministry of Education, Science and Technology and the World Class University program of NRF, Republic of Korea; the Lithuanian Academy of Sciences; the Ministry of Education, and University of Malaya (Malaysia); the Mexican Funding Agencies (CINVESTAV, CONACYT, SEP, and UASLP-FAI); the Ministry of Business, Innovation and Employment, New Zealand; the Pakistan Atomic Energy Commission; the Ministry of Science and Higher Education and the National Science Centre, Poland; the Funda\c{c}\~ao para a Ci\^encia e a Tecnologia, Portugal; JINR, Dubna; the Ministry of Education and Science of the Russian Federation, the Federal Agency of Atomic Energy of the Russian Federation, Russian Academy of Sciences, and the Russian Foundation for Basic Research; the Ministry of Education, Science and Technological Development of Serbia; the Secretar\'{\i}a de Estado de Investigaci\'on, Desarrollo e Innovaci\'on and Programa Consolider-Ingenio 2010, Spain; the Swiss Funding Agencies (ETH Board, ETH Zurich, PSI, SNF, UniZH, Canton Zurich, and SER); the Ministry of Science and Technology, Taipei; the Thailand Center of Excellence in Physics, the Institute for the Promotion of Teaching Science and Technology of Thailand, Special Task Force for Activating Research and the National Science and Technology Development Agency of Thailand; the Scientific and Technical Research Council of Turkey, and Turkish Atomic Energy Authority; the National Academy of Sciences of Ukraine, and State Fund for Fundamental Researches, Ukraine; the Science and Technology Facilities Council, UK; the US Department of Energy, and the US National Science Foundation.

Individuals have received support from the Marie-Curie programme and the European Research Council and EPLANET (European Union); the Leventis Foundation; the A. P. Sloan Foundation; the Alexander von Humboldt Foundation; the Belgian Federal Science Policy Office; the Fonds pour la Formation \`a la Recherche dans l'Industrie et dans l'Agriculture (FRIA-Belgium); the Agentschap voor Innovatie door Wetenschap en Technologie (IWT-Belgium); the Ministry of Education, Youth and Sports (MEYS) of the Czech Republic; the Council of Science and Industrial Research, India; the HOMING PLUS programme of Foundation for Polish Science, cofinanced from European Union, Regional Development Fund; the Compagnia di San Paolo (Torino); the Consorzio per la Fisica (Trieste); MIUR project 20108T4XTM (Italy); the Thalis and Aristeia programmes cofinanced by EU-ESF and the Greek NSRF; and the National Priorities Research Program by Qatar National Research Fund.
\end{acknowledgments}
\bibliography{auto_generated}

\cleardoublepage \appendix\section{The CMS Collaboration \label{app:collab}}\begin{sloppypar}\hyphenpenalty=5000\widowpenalty=500\clubpenalty=5000\textbf{Yerevan Physics Institute,  Yerevan,  Armenia}\\*[0pt]
V.~Khachatryan, A.M.~Sirunyan, A.~Tumasyan
\vskip\cmsinstskip
\textbf{Institut f\"{u}r Hochenergiephysik der OeAW,  Wien,  Austria}\\*[0pt]
W.~Adam, T.~Bergauer, M.~Dragicevic, J.~Er\"{o}, C.~Fabjan\cmsAuthorMark{1}, M.~Friedl, R.~Fr\"{u}hwirth\cmsAuthorMark{1}, V.M.~Ghete, C.~Hartl, N.~H\"{o}rmann, J.~Hrubec, M.~Jeitler\cmsAuthorMark{1}, W.~Kiesenhofer, V.~Kn\"{u}nz, M.~Krammer\cmsAuthorMark{1}, I.~Kr\"{a}tschmer, D.~Liko, I.~Mikulec, D.~Rabady\cmsAuthorMark{2}, B.~Rahbaran, H.~Rohringer, R.~Sch\"{o}fbeck, J.~Strauss, A.~Taurok, W.~Treberer-Treberspurg, W.~Waltenberger, C.-E.~Wulz\cmsAuthorMark{1}
% ...
\vskip\cmsinstskip
% ...
1:~~Also at Vienna University of Technology, Vienna, Austria\\
2:~~Also at CERN, European Organization for Nuclear Research, Geneva, Switzerland\\
% ...

\end{sloppypar}
\end{document}